\documentclass[reprint,aip,jcp,amsmath]{revtex4-1}

\usepackage{graphicx}
\usepackage{color}

\newcommand{\Ket}[1]{\vert#1\rangle}
\newcommand{\Bra}[1]{\langle#1\vert}
\newcommand{\T}[1]{\mathrm{#1}}
\newcommand{\V}[1]{\boldsymbol{#1}}
\newcommand{\Op}[1]{\hat{#1}}
\newcommand{\BS}[1]{({#1})}

\begin{document}

\title{Vibronic exciton theory of singlet fission.~I.~Linear absorption and the anatomy of the correlated triplet pair state}

\author{Roel Tempelaar}
\email{r.tempelaar@gmail.com}
\author{David R.~Reichman}
\email{drr2103@columbia.edu}
\affiliation{Department of Chemistry, Columbia University, 3000 Broadway, New York, New York 10027, USA}

\begin{abstract}
Recent time-resolved spectroscopic experiments have indicated that vibronic coupling plays a vital role in facilitating the process of singlet fission. In this work, which forms the first article of a series, we set out to unravel the mechanisms underlying singlet fission through a vibronic exciton theory. We formulate a model in which both electronic and vibrational degrees of freedom are treated microscopically and non-perturbatively. Using pentacene as a prototypical material for singlet fission, we subject our theory to comparison with measurements on polarization-resolved absorption of single crystals, and employ our model to characterize the excited states underlying the absorption band. Special attention is given to convergence of photophysical observables with respect to the basis size employed, through which we determine the optimal basis for more expensive calculations to be presented in subsequent work. We furthermore evaluate the energetic separation between the optically prepared singlet excited state and the correlated triplet pair state, as well as provide a real-space characterization of the latter, both of which are of key importance in the discussion of fission dynamics. We discuss our results in the context of recent experimental studies.
\end{abstract}

\maketitle

\section{Introduction}

Singlet fission, the molecular process whereby a singlet excited state is converted into two triplet excitons, has been discussed in the literature for over five decades.\cite{Singh_65, Avakian_68a}  In recent years, a renewed focus on the topic has emerged, driven by both fundamental interest and the promise of improved photovoltaic devices.\cite{Smith_10a, Smith_13a} With respect to the latter, the splitting of the absorbed photon energy in two lower-energy packets enables the possibility of circumventing\cite{Hanna_06a} the Shockley-Queisser efficiency limit\cite{Shockley_61a} for single-junction photovoltaic materials. Furthermore, the triplet excitons that are produced in the process are stable against radiative losses, having no dipole-allowed optical transitions to the (singlet) ground state.  This property facilitates a much longer range of energy transport towards extraction regions than would be possible with singlet excitons.\cite{Najafov_10a, Irkhin_11a} The technological potential of singlet fission has been demonstrated in various proof-of-principle studies using polyacenes as a fission material.\cite{Jadhav_11a, Ehler_12a, Congreve_13a, Yang_15a} Nevertheless, beyond polyacenes, singlet fission has remained a somewhat rare and relatively exotic phenomenon, and a much-needed expansion of the spectrum of suitable materials is hampered by a lack of microscopic understanding of singlet fission through which the determining factors of a complete set of fission materials can be identified.\cite{Smith_13a}

An important question surrounding singlet fission that has remained incompletely answered is the surprisingly fast timescale of exciton multiplication in certain acene systems such as bulk pentacene. In such systems efficient fission is found to occur very rapidly,\cite{Wilson_11a, Chan_11a} and evidence exists that a near instantaneous population of crucial fission intermediates concomitantly occurs.\cite{Chan_11a, Bakulin_16a, Monahan_16a} Early attempts to account for these observations relied on very strong electronic coupling between the initially excited singlet exciton and the correlated triplet pair state,\cite{Chan_11a} the fission intermediate which renders the overall process spin-allowed. However such sizable coupling seems inconsistent with most electronic structure calculations.\cite{Zimmerman_11a, Feng_13a, Parker_14a} Recent time-resolved spectroscopic studies, seeking to find alternative explanations, have pointed to the importance of vibronic coupling.\cite{Musser_15a, Bakulin_16a, Monahan_16a} Transient absorption measurements on pentacene derivatives suggested that a conical intersection is the driving mechanism for fission.\cite{Musser_15a} Yet another scenario for pentacene was proposed based on two-dimensional electronic spectroscopy (2DES), in which high-frequency intramolecular vibrations generate a resonance between the singlet and correlated triplet pair states by matching their energy difference.\cite{Bakulin_16a} Two-photon photoemission measurements on hexacene were interpreted similarly,\cite{Monahan_16a} despite the very different energetics of this material.

Given the recent indications of the importance of vibronic coupling to singlet fission, a theoretical study of fission materials is called for in which the involved electronic degrees of freedom and interacting intramolecular modes are treated microscopically and non-perturbatively. The aforementioned spectroscopic studies were mostly supported by calculations performed using phenomenological models, including a minimal number of electronic degrees of freedom.\cite{Chan_11a, Bakulin_16a, Monahan_16a} Although such modeling is extremely helpful for interpreting the spectroscopic measurements, it does not provide fully microscopic insights at the molecular level. On the other hand, available high-level theoretical studies on fission materials\cite{Zimmerman_10a, Zimmerman_11a, Havenith_12a, Feng_13a, Casanova_14a} are numerically constrained to identifying excited states in a time-independent framework for limited cluster sizes. A viable approach to bridge the gap between these calculations and measurements of singlet fission is through microscopic modeling. Such has been proven successful in several studies on fission materials,\cite{Yamagata_11a, Teichen_12a, Beljonne_13a, Berkelbach_13a, Berkelbach_13b, Berkelbach_14a, Hestand_15a} however, applications did not address the fission dynamics,\cite{Yamagata_11a, Beljonne_13a, Hestand_15a} or were limited to a perturbative treatment of electronic couplings \cite{Teichen_12a} or the vibrational degrees of freedom.\cite{Berkelbach_13a, Berkelbach_13b, Berkelbach_14a}

The present article forms the first entry in a series of studies in which we present a microscopic vibronic exciton model aimed at unraveling the mechanistic principles underlying singlet fission, with a particular emphasis placed on the role of vibronic coupling. We provide a detailed description of the model in which both the electronic degrees of freedom and a selected intramolecular vibrational mode are treated non-perturbatively. In part, the applied methodology descends from a recent parametrization of polarization-resolved absorption of single-crystalline pentacene by Hestand \textit{et al.},\cite{Hestand_15a} which we adopt and combine with a treatment of the relevant triplet excited states not previously considered.\cite{Hestand_15a} In doing so, we base our parametrization on the first direct detection of the correlated triplet pair state through the aforementioned recent 2DES experiments on pentacene.\cite{Bakulin_16a} We will further investigate this measurement in a follow-up companion study focused on 2DES of fission materials.  In order to realize (expensive) 2DES calculations, we address in the present article the convergence of the photophysical observables with respect to the crystal dimensions, from which we determine the optimal crystal size to simulate the photophysics of fission materials on a cost-effective yet accurate basis. This will also pave the way for the final piece in this series, a study of the dynamical evolution of the full singlet fission process by marrying our model with quantum dynamical methods, which forms a critical test for the functional mechanisms hypothesized to underly this process.\footnote{After completion of our manuscript and during the formulation of our dynamical modeling, we became aware of Y.~Fujihashi, L.~Chen, A.~Ishizaki, J.~Wang, and Y.~Zhao, J.~Chem.~Phys.~\textbf{146}, 044101 (2017), which non-perturbatively explores the effect of high-frequency modes on singlet fission using dynamical calculations based on a phenomenological model. A comparison of this work with the dynamics that emerge from the microscopic Hamiltonian will be made in the third paper of this series.}

In the present article, we use linear absorption of crystalline pentacene as a means to determine the accuracy of the proposed model. At the same time, the model allows us to characterize the excited states underlying the absorption band, through which we shed light on the energetic separation between the singlet exciton and the correlated triplet pair state. In particular, we demonstrate how this separation is dependent on the applied crystal sizes. Another topic that we address is the spatial configuration of the correlated triplet pair state, which has come under debate with a recently reported triplet-triplet separation of several times the intermolecular distance\cite{Wang_15a} contesting the conventional idea of triplets located at neighboring molecules.\cite{Smith_10a, Smith_13a} The microscopic nature of our model provides direct access to the anatomy of this elusive state, which helps to shed light on this matter.

The paper is organized as follows. In Sec.~\ref{Sec_Theory}, we present the vibronic basis set and Hamiltonian, followed by a discussion of the associated parameters for pentacene. Results for pentacene are presented and discussed in Sec.~\ref{Sec_Results}. First, the linear absorption spectrum is analyzed, highlighting the composition of the underlying excited states. What follows is an evaluation of the excited state energies, and other photophysical properties, in relation to the applied crystal size. Finally, we analyze the spatial composition of the correlated triplet pair state. We conclude in Sec.~\ref{Sec_Conclusions}.

\section{Theory}\label{Sec_Theory}

Before presenting our theoretical framework, we begin with a few remarks on the distinction between adiabatic and diabatic excited states. Adiabatic states (eigenstates of the Hamiltonian of the fission material) belong to the quantum basis that is accessible through spectroscopy. Important examples of such states in the context of singlet fission are the optically prepared singlet excited state S$_1$, and the subsequently populated correlated triplet pair state TT$_1$, where it is understood that both labels in reality might represent a manifold of energetically closely-spaced adiabatic states. Nevertheless, we instead formulate our model in the diabatic picture (also referred to as site basis or molecular basis). The underlying approximations, as well as the utility of the diabatic representation to describe singlet fission, are extensively discussed in Ref.~\citenum{Berkelbach_13a}. From our diabatic calculations, the adiabatic spectroscopic observables are readily extracted. In order to clearly distinguish between these two bases, we will consistently denote diabatic states with lower case labels, such as s$_1$ and t$_1$. On the other hand, it should be understood that adiabatic states such as S$_1$ and TT$_1$ are never composed entirely of s$_1$ and t$_1$, respectively. Rather, they consist of some admixture of the two, additionally mixed with other states such as diabatic charge transfer states.\cite{Yamagata_11a, Beljonne_13a, Berkelbach_14a, Hestand_15a} In this sense, the adiabatic states are labeled so as to distinguish between a predominant singlet or triplet composition.

\subsection{Basis set}\label{Sec_Basis}

Similarly to Refs.~\citenum{Teichen_12a, Yamagata_11a, Beljonne_13a, Berkelbach_13a, Berkelbach_13b, Berkelbach_14a, Hestand_15a}, our diabatic basis comprises the many-body states formed by excitations within the minimal active space of all highest and lowest unoccupied molecular orbitals (HOMOs and LUMOs) of the \textit{isolated} molecules. Explicit formulations of the symmetry-adapted linear combinations of the diabatic basis states can be found in Ref.~\citenum{Berkelbach_13a}. In addition to the ground state s$_0$, in which the HOMO is fully occupied, we include for each molecule the aforementioned singlet excitations through the basis states $\Ket{\BS{\T{s}_1}_m}$. Here, an electron is promoted to the LUMO on a molecule labeled $m$, leaving a hole in the corresponding HOMO, and with all other molecules in their respective s$_0$ state. We also consider charge transfer (CT) states, for which a hole (cation) and electron (anion) are found at different molecules. These states are denoted $\Ket{\BS{\T{c}}_m,\BS{\T{a}}_{m'}}$, with $m$ and $m'$ labeling the molecular sites of the cation and anion, respectively. Likewise, the correlated triplet pair state is accounted for by the basis states $\Ket{\BS{\T{t}_1}_m,\BS{\T{t}_1}_{m'}}$, where two triplets located at molecules $m$ and $m'$, respectively, are entangled in an overall spin-zero state.

In addition to the electronic degrees of freedom, we explicitly include a single intramolecular vibrational mode in the basis set. Such an approach has been employed in earlier microscopic studies on fission materials,\cite{Yamagata_11a, Beljonne_13a, Hestand_15a} although there is only a single report in which the basis encompassed the triplet states relevant to fission.\cite{Beljonne_13a} We note that multiple modes have been explicitly included in the phenomenological model used to support the recent 2DES study on singlet fission in pentacene.\cite{Bakulin_16a} Although such an extension of the basis set is straightforward in principle, it becomes numerically prohibitive at the microscopic level employed in this work. We therefore limit ourselves to the explicit treatment of the spectrally most dominant mode, as was done in the aforementioned microscopic studies.\cite{Yamagata_11a, Beljonne_13a, Hestand_15a} We do bear in mind, however, that such an approach can easily be combined with a Redfield-type propagation scheme in which the remainder of the vibrations are treated perturbatively.

A full quantum treatment of the vibrational mode, in principle, entails a basis set that scales very unfavorably with increasing number of molecules. Fortunately, previous studies have demonstrated that the photophysics of fission materials is typically well accounted for when limiting the basis to so-called ``two-particle'' states,\cite{Yamagata_11a, Beljonne_13a, Hestand_15a} in which the eigenstates are decomposed into sums of pairwise (electronic and/or vibrational) excitations, an approximation that dramatically reduces the relevant Hilbert space.\cite{Philpott_71a, Spano_02a} To keep computations manageable, we therefore restrict ourselves to the two-particle basis. For the ease of discussion, we proceed to formulate our theory in this basis, although noting that a general formulation in the multi-particle basis is straightforward. Accordingly, we formulate (adiabatic) eigenstates of the Hamiltonian of the material as
\begin{align}
\label{Eq_Eig}
\Ket{\alpha}=&\;\sum_{m,\tilde{\nu}}c^{\T{s}_1}_{\alpha;m,\tilde{\nu}}\Ket{\BS{\T{s}_1}_m,\tilde{\nu}}\\
&+\sum_{m\neq m'}\sum_{\tilde{\nu}}\sum_{\nu'\geq1}c^{\T{s}_1\T{s}_0}_{\alpha;m,\tilde{\nu},m',\nu'}\Ket{\BS{\T{s}_1}_m,\tilde{\nu},\BS{\T{s}_0}_{m'},\nu'}\nonumber\\
&+\sum_{m\neq m'}\sum_{\nu_+,\nu_-}c^{\T{ca}}_{\alpha;m,\nu_+,m',\nu_-}\Ket{\BS{\T{c}}_m,\nu_+,\BS{\T{a}}_{m'},\nu_-}\nonumber\\
&+\sum_{m>m'}\sum_{\bar\nu,\bar\nu'}c^{\T{t}_1\T{t}_1}_{\alpha;m,\bar\nu,m',\bar\nu'}\Ket{\BS{\T{t}_1}_m,\bar\nu,\BS{\T{t}_1}_{m'},\bar\nu'}.\nonumber
\end{align}
Here, the first summation extends over the vibronic basis states involving the electronic excitation $\Ket{\BS{\T{s}_1}_m}$ accompanied by $\tilde{\nu}$ vibrational quanta in the associated nuclear potential, while $\Ket{\BS{\T{s}_1}_m,\tilde{\nu}}$ is a short-hand notation for $\Ket{\BS{\T{s}_1}_m}\otimes\Ket{\tilde{\nu}}$. In a similar fashion, the second summand contains basis states for which such a vibronic excitation is accompanied by a purely vibrational excitation involving $\nu'(\geq1)$ quanta in the s$_0$ potential located at molecule $m'$. The subsequent term contains the CT states with $\nu_+$ and $\nu_-$ quanta in the cationic and anionic potentials, respectively. The final summation extends over the triplet pairs, where the associated vibrational quanta $\bar\nu$ and $\bar\nu'$ refer to the t$_1$ vibrational potential. Here, the summation is restricted to $m>m'$ in order to avoid double counting.

\subsection{Hamiltonian}\label{Sec_Hamiltonian}

The expansion coefficients in Eq.~\ref{Eq_Eig} are obtained upon numerically solving the eigenvalue equation, $\Op{H}\Ket{\alpha}=\omega_\alpha\Ket{\alpha}$, where $\Op{H}$ is the Hamiltonian, and $\omega_\alpha$ represents the eigenenergy associated with state $\alpha$ ($\hbar=1$ is assumed throughout this paper). For clarity, we subdivide the Hamiltonian into parts,\cite{Beljonne_13a}
\begin{align}
\Op{H}=&\;\Op{H}_{\T{s}_1}+\Op{H}_{\T{s}_1-\T{ca}}+\Op{H}_{\T{ca}}+\Op{H}_{\T{ca}-\T{t}_1\T{t}_1}+\Op{H}_{\T{t}_1\T{t}_1}\nonumber\\
&+\Op{H}_{\omega_0}+\Op{H}_{\lambda},
\label{Eq_HamTotal}
\end{align}
each associated with different physical mechanisms. Note that no contribution associated with the electronic ground state s$_0$ is included here, implying that this state is associated with the zero point of energy.

The first part of the Hamiltonian is given by
\begin{align}
\Op{H}_{\T{s}_1}=&\;E_{\T{s}_1}\sum_{m}\Ket{\BS{\T{s}_1}_m}\Bra{\BS{\T{s}_1}_m}\nonumber\\
&+\sum_{m\neq m'}J_{m,m'}\Ket{\BS{\T{s}_1}_m}\Bra{\BS{\T{s}_1}_{m'}},
\end{align}
and accounts for the diagonal energies of the diabatic $\T{s}_1$ states, denoted as $E_{\T{s}_1}$ and taken to be equal for all molecules. This part also describes dipole-dipole interactions between these states, denoted as $J_{m,m'}$.

The second part describes the dissociation of the Frenkel-type $\T{s}_1$ states into CT excitons,
\begin{align}
\Op{H}_{\T{s}_1-\T{ca}}=&\;\sum_{m\neq m'}t^\T{HH}_{m,m'}\Ket{\BS{\T{s}_1}_m}\Bra{\BS{\T{c}}_{m'},\BS{\T{a}}_m}\\
&+\sum_{m\neq m'}t^\T{LL}_{m,m'}\Ket{\BS{\T{s}_1}_m}\Bra{\BS{\T{c}}_m,\BS{\T{a}}_{m'}}+\T{H.c.},\nonumber
\end{align}
which is mediated by the HOMO-HOMO and LUMO-LUMO charge overlap integrals, $t^\T{HH}_{m,m'}$ and $t^\T{LL}_{m,m'}$, transferring electrons and holes, respectively.

The third part is given by,
\begin{align}
\Op{H}_{\T{ca}}=&\;\sum_{m\neq m'}(U_{m,m'}+E_{\T{s}_1})\Ket{\BS{\T{c}}_m,\BS{\T{a}}_{m'}}\Bra{\BS{\T{c}}_m,\BS{\T{a}}_{m'}}\nonumber\\
&+\sum_{m\neq m'\neq m''}t^\T{HH}_{m,m'}\Ket{\BS{\T{c}}_m,\BS{\T{a}}_{m''}}\Bra{\BS{\T{c}}_{m'},\BS{\T{a}}_{m''}}\\
&+\sum_{m\neq m'\neq m''}t^\T{LL}_{m,m'}\Ket{\BS{\T{c}}_{m''},\BS{\T{a}}_{m}}\Bra{\BS{\T{c}}_{m''},\BS{\T{a}}_{m'}}\nonumber,
\end{align}
and accounts for separated charges at molecules $m$ and $m'$, whose Coulomb energy relative to the diabatic energy of s$_1$ is given by $U_{m,m'}$, as well as the short-ranged CT interactions mediated by the charge overlap integrals.

The next two parts are associated with the triplet pair excitations. The first one couples such states to
CT states,
\begin{align}
\Op{H}_{\T{ca}-\T{t}_1\T{t}_1}=\sqrt{\frac{3}{2}}\Big(\;\sum_{m\neq m'}t^\T{HL}_{m,m'}\Ket{\BS{\T{a}}_m,\BS{\T{c}}_{m'}}\Bra{\BS{\T{t}_1}_{m}\BS{\T{t}_1}_{m'}}\nonumber\\
+\sum_{m\neq m'}t^\T{LH}_{m,m'}\Ket{\BS{\T{c}}_m,\BS{\T{a}}_{m'}}\Bra{\BS{\T{t}_1}_{m}\BS{\T{t}_1}_{m'}}\Big)+\T{H.c.},
\end{align}
where $t^\T{HL}_{m,m'}$ denotes the charge overlap integral which transfers an electron from the HOMO of molecule $m$ to the LUMO of molecule $m'$, and oppositely for $t^\T{LH}_{m,m'}$. The second part describes the diagonal energies of the triplet pairs,
\begin{align}
\Op{H}_{\T{t}_1\T{t}_1}=E_{\T{t}_1\T{t}_1}\sum_{m>m'}\Ket{\BS{\T{t}_1}_m,\BS{\T{t}_1}_{m'}}\Bra{\BS{\T{t}_1}_m,\BS{\T{t}_1}_{m'}},
\end{align}
quantified by $E_{\T{t}_1\T{t}_1}$, and again assumed site-independent.

The last two parts of the Hamiltonian relate to the intramolecular vibrational mode. Assuming a harmonic potential, the total vibrational energy is given by
\begin{align}
\Op{H}_{\omega_0}=\omega_0\sum_m\Op{b}_m^\dagger\Op{b}_m,
\end{align}
where $\omega_0$ is the vibrational quantum ($\hbar=1$), and $\Op{b}_m^{(\dagger)}$ is the annihilation (creation) operator pertaining to the s$_0$ potential at molecule $m$. This mode is coupled linearly to the electronic diagonal energies (Holstein-type interaction), through
\begin{align}
\Op{H}_{\lambda}=\omega_0\!\!\!\sum_{x=\T{s}_1,\T{c},\T{a},\T{t}_1}\sum_m\;\lambda_x(\Op{b}_m^\dagger+\Op{b}_m+\lambda_x)\Ket{\BS{x}_m}\Bra{\BS{x}_m}.
\end{align}
Here, $x$ runs over all possible diabatic states associated with each molecule. $\lambda_x$ represents the shift of the vibrational equilibrium position for each state, using the s$_0$ potential as a reference. The square of the vibrational displacements relative to the ground state, $\lambda_x^2$, yields the associated Huang-Rhys (HR) factor. Since in our approach all vibrational states are represented in their respective eigenbasis, the electronic interaction terms contributing to the Hamiltonian results in a mixing between such states, dictated by vibrational overlap factors.\cite{Spano_09a}

Extension of Eq.~\ref{Eq_HamTotal} to include two-electron couplings is straightforward (and essentially invokes a direct coupling between the singlet excited states and triplet pairs, $\Op{H}_{\T{s}_1-\T{t}_1\T{t}_1}$). Nevertheless, their magnitudes are known to be an order of magnitude smaller than the one-electron integrals discussed above,\cite{Zimmerman_11a, Havenith_12a, Berkelbach_13b, Chan_13a, Smith_13a} and a significant contribution of such couplings to fission dynamics is therefore unlikely, at least in systems like pentacene.\cite{Berkelbach_13b} We have performed additional calculations including typical values for the two-electron couplings, and do not find noteworthy differences for the results reported in this work. We therefore disregard such couplings for the sake of simplicity. However, a brief discussion of the significance of two-electron couplings between local triplet excitations is presented in Sec.~\ref{Sec_TripletPair}.

\subsection{Parameters for pentacene}\label{Sec_Parms}

Pentacene arguably is the most widely studied fission material, attracting great interest due to its nearly unrivaled fission rate and high fission efficiency. At the same time, the rapid and efficient fission dynamics found for this material poses a challenge for theoretical modeling. These factors, as well as the wealth of information known about this material, motivate us to use pentacene as an exemplary singlet fission system in this series of articles. We expect many of our conclusions to be general for other strongly coupled acene crystals. The parameters for pentacene used in our calculations are to a large extent identical to those reported by Hestand \textit{et al.},\cite{Hestand_15a} which were obtained through a combination of quantum chemical calculation and fitting to polarized absorption spectra of pentacene single crystals.\cite{Beljonne_13a, Hestand_15a} We proceed with a detailed discussion of these parameters, in particular highlighting the adaptations made in our modeling. We note the qualitatively, and in some cases semi-quantitative, agreement with the parameters obtained via a similar approach by Berkelbach \textit{et al.}~for pentacene dimers\cite{Berkelbach_13b} and single crystals.\cite{Berkelbach_14a} This similarity gives confidence in the robustness of the general philosophy of our work and these past studies.

\begin{figure}
\includegraphics[width=8.5cm]{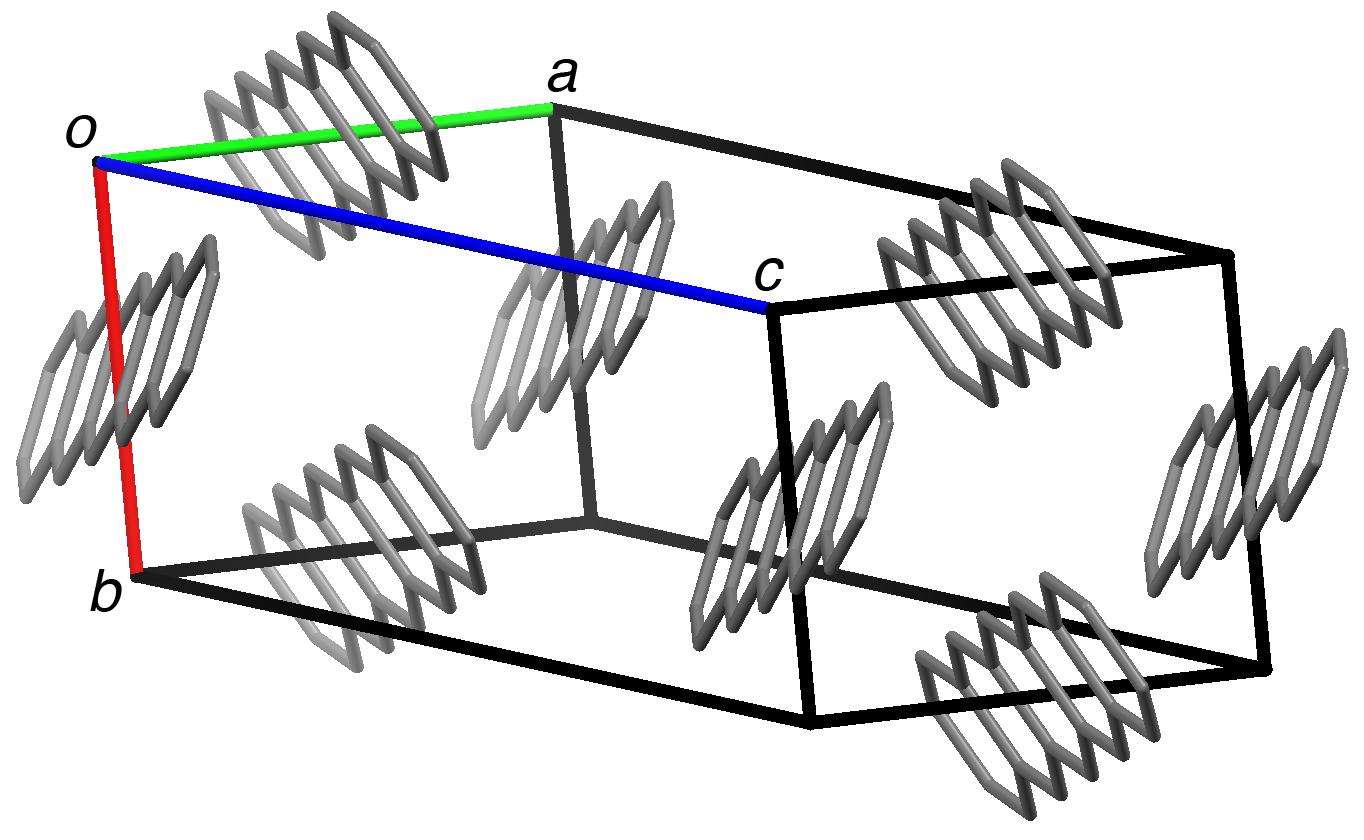}
\caption{Crystal structure of pentacene, taken from Ref.~\citenum{Holmes_99a}. The crystallographic $a$-, $b$-, and $c$-axis are indicated with green, red, and blue, respectively. The unit cell contains two inequivalent molecules with their long molecular axes oriented roughly along $c$, while their short axes form a characteristic herringbone structure in the $ab$-plane.}
\label{Fig_Crystal}
\end{figure}

In the crystal phase, pentacene molecules form a triclinic system with two inequivalent molecules per unit cell.\cite{Holmes_99a, Mattheus_01a} This is illustrated in Fig.~\ref{Fig_Crystal}, showing the crystal structure taken from Ref.~\citenum{Holmes_99a}, including the crystallographic axes. As can be seen, the long axes of the pentacene molecules are oriented roughly along the $c$-axis, while the short axes form the characteristic herringbone structure in the $ab$-plane. Similarly to Hestand \textit{et al.},\cite{Hestand_15a} we consider in our calculations a herringbone lattice consisting of $M\times M$ unit cells extending in the $ab$-plane, which corresponds to $2M^2$ molecules (such that $m$ runs from 1 to $2M^2$). The lengths of the $a$ and $b$ lattice constants are found to be 7.71 and 6.28~\AA, respectively, with a mutual angle of 84.5$^\circ$, while the separation between the inequivalent molecules within the same cell are is given by $(a,b)=(1/2,1/2)$.\cite{Holmes_99a} The molecular positions are determined based on this structural data by taking for each molecule its point of inversion symmetry.

\begin{table}
\begin{center}
\begin{tabular}{l l l l}
\hline
\hline
Parameter & Symbol & Value & \\
\hline
s$_0-$s$_1$ energy & $E_{\T{s}_1}$ & 2.09~eV & (16\;890~cm$^{-1}$) \\
s$_0-$t$_1$t$_1$ energy & $E_{\T{t}_1\T{t}_1}$ & 1.75~eV & (14\;140~cm$^{-1}$) \\
Vibr.~energy & $\omega_0$ & 0.17~eV & (1380~cm$^{-1}$) \\
Huang-Rhys factors & & & \\
\quad s$_0-$s$_1$ & $\lambda_{\T{s}_1}^2$ & 1.1 & \\
\quad s$_0-$a & $\lambda_{\T{a}}^2$ & 0.29 & \\
\quad s$_0-$c & $\lambda_{\T{c}}^2$ & 0.39 & \\
\quad s$_0-$t$_1$ & $\lambda_{\T{t}_1}^2$ & 1.1 & \\
\hline
\hline
\end{tabular}
\caption{Parameters applied in our model. For electronic couplings, see the supplementary material.}
\label{Tab_Parms}
\end{center}
\end{table}

The most important parameters applied in our model are summarized in Tab.~\ref{Tab_Parms}, while the electronic couplings are presented in the supplementary material. The dipole-dipole couplings $J_{m,m'}$ in crystalline pentacene are rather weak, and have therefore been neglected by Berkelbach \textit{et al.}\cite{Berkelbach_13b, Berkelbach_14a} Here, we use values reported by Hestand \textit{et al.},\cite{Hestand_15a} which have been calculated through INDO/CCSD using the coordinates from Ref.~\citenum{Holmes_99a}, and screened (divided) by a directionally averaged optical dielectric constant $\epsilon=3.5$.\cite{Tsiper_03a}

The applied one-electron HOMO-HOMO and LUMO-LUMO couplings, $t^\T{HH}_{m,m'}$ and $t^\T{LL}_{m,m'}$, were originally reported in Ref.~\citenum{Beljonne_13a} based on density functional theory (using B3LYP and a double zeta basis set). Similarly to Hestand \textit{et al.},\cite{Hestand_15a} we apply a scaling factor of 1.1 in order to reproduce the Davydov splitting observed for pentacene single crystals. We further note that Berkelbach \textit{et al.}~found somewhat larger, yet qualitatively similar couplings through \textit{ab initio} calculations using the Hartree-Fock molecular orbitals of isolated pentacene molecules and a 6-31G(d) basis set.\cite{Berkelbach_13b} However, a dramatic down-scaling of these couplings was found to be necessary in order to reproduce the Davydov components.\cite{Berkelbach_14a} We point out that such is not necessary once vibronic coupling is accounted for, which by itself has a strong mitigating impact on the Davydov splitting (see for example Fig.~5 from Ref.~\citenum{Hestand_15a}, where a direct comparison is drawn between vibronic and purely-electronic calculations).

The model applied by Hestand \textit{et al.}\cite{Hestand_15a}~did not include triplet states, hence at this point our modeling and parameters start to differ from those used in that work. For the HOMO-LUMO couplings $t^\T{HL}_{m,m'}$, not reported by Hestand \textit{et al.}, we adapt the couplings calculated by Berkelbach \textit{et al.}\cite{Berkelbach_13b} Similarly to their work on crystalline pentacene,\cite{Berkelbach_14a} we apply an overall rescaling of these couplings, while keeping the signs and relative differences among them unaltered. In order to find an appropriate scaling factor, we have determined the average magnitude of the HOMO-HOMO and LUMO-LUMO couplings of Berkelbach \textit{et al.}~relative to those by Hestand \textit{et al.}~(found to be 0.68), and rescaled the HOMO-LUMO couplings accordingly.

By employing the vibronic basis set outlined in Sec.~\ref{Sec_Basis}, we provide an exact treatment of the symmetric stretching vibration with a frequency $\omega_0=0.17$~eV (1380~cm$^{-1}$) which is the dominant progression building mode observed in absorption spectra of both dissolved and crystalline pentacene.\cite{Yamagata_11a, Beljonne_13a, Hestand_15a} Its HR factor associated with the diabatic singlet state is taken to be $\lambda_{\T{s}_1}^2=1.1$, a value derived from fitting the vibronic peak areas of dissolved pentacene to a Poisson distribution.\cite{Hestand_15a} The ionic HR factors ($\lambda_\T{a}^2=0.29$ and $\lambda_\T{c}^2=0.39$)\cite{Hestand_15a} are based on calculations reported in Ref.~\citenum{Coropceanu_07a}. We note that these values are significantly smaller than the singlet state HR factor, something that has also been observed in terrylene crystals.\cite{Yamagata_14a, Hestand_15b}. Relatively little is known about the HR factor pertaining to t$_1$. In Ref.~\citenum{Beljonne_13a}, the weak dependence of linear absorption on $\lambda_{\T{t}_1}^2$ was noted and an \textit{ad hoc} value of 1.0 was applied. In contrast, a value of 0 was used in the phenomenological modeling reported in Ref.~\citenum{Bakulin_16a}, based on a fitting to 2DES data. However, recent calculations on tetracene have shown the effective HR factor of triplet pair states to be competitive with $\lambda_{\T{s}_1}^2$.\cite{Ito_15a} Furthermore, vibrational overtones associated with $\omega_0$ observed in S$_0-$T$_1$ absorption of anthracene and its derivates using the heavy-atom effect suggests this factor to be substantial.\cite{McGlynn_64a} Expecting the HR factors of acenes to be reasonably similar, we therefore set $\lambda_{\T{t}_1}^2=1.1$.

The diagonal energies appearing in the Hamiltonian are notoriously difficult to quantitatively determine based on unaltered first-principles quantum chemical methods (see also the discussion in Ref.~\citenum{Berkelbach_13a}). Most microscopic models for fission materials have instead used values taken from experiments, mostly through fitting to linear absorption.\cite{Yamagata_11a, Beljonne_13a, Berkelbach_14a, Hestand_15a} For example, such fitting was applied in Ref.~\citenum{Beljonne_13a} to determine the Coulomb energies $U_{m,m'}$ of crystalline pentacene. Here we use the slightly modified values proposed by Hestand \textit{et al.}\cite{Hestand_15a} (which are summarized in the supplementary material). Regarding the diabatic $\T{s}_0\rightarrow\T{s}_1$ transition energy, our value is slightly red-shifted relative to the one reported in that work,\cite{Hestand_15a} namely $E_{\T{s}_1}=2.09$~eV (16\;890~cm$^{-1}$). Still, this value is higher than the energy used by Berkelbach \textit{et al.}\cite{Berkelbach_14a}~by about 100 meV. The origin of this discrepancy is discussed in Sec.~\ref{Sec_RedShift}. Lastly, the energy of the diabatic triplet pair excitation is taken to be $E_{\T{t}_1\T{t}_1}=1.75$~eV (14\;140~cm$^{-1}$), which yields adiabatic TT$_1$ state energies in agreement with 2DES measurements\cite{Bakulin_16a} as discussed in Sec.~\ref{Sec_RedShift}.

In order to keep calculations computationally manageable, the basis set presented in Sec.~\ref{Sec_Basis} is minimized as much as possible while retaining convergence for the photophysics of interest. Similarly to Hestand \textit{et al.},\cite{Hestand_15a} the maximum charge separation is set to 20~\AA~(where the separation is based on the molecular positions). Instead of also adapting this value for the maximum separation for non-ionic two-particle states,\cite{Hestand_15a} we have found that a smaller truncation radius of 7~\AA~suffices. The physical origin of such a truncation lies in the finite extent of the vibrational distortion surrounding the electronic excitation.\cite{Tempelaar_13a} We furthermore reduced the total number of vibrational quanta per basis state to 2. The maximum separation for the triplet pair excitations was limited to 6.2~\AA, such that only nearest-neighbors in the $(a,b)=(1/2,1/2)$ and $(a,b)=(1/2,-1/2)$ directions contribute. The case of a basis with a more extended set of pair states is explored in Sec.~\ref{Sec_TripletPair}. Lastly, periodic boundary conditions are imposed. Accordingly, the coupling between two molecules is determined based on the minimal real-space separation of these molecules resulting from any translation of one molecule by multiples of $M$ lattice spacings in the $a$- and $b$-directions.

\section{Results and discussion}\label{Sec_Results}

\subsection{Linear absorption and excited state admixtures}\label{Sec_LinAbs}

Earlier studies of the theory of singlet fission have employed the comparison to experimental linear absorption as a means of obtaining parameters and as a critical test of the modeling itself.\cite{Yamagata_11a, Beljonne_13a, Berkelbach_14a, Hestand_15a} Such an approach has proven to be particularly fruitful when applied to polarization-resolved spectra of single crystals, which provides detailed insight into the involved excited states.\cite{Berkelbach_13a, Hestand_15a} The theory presented here is partly derived from earlier work on pentacene which centered on such a comparison.\cite{Hestand_15a} In a follow-up article, we apply an additional stringent test via the comparison to 2DES measurements. Nonetheless, we will start by revisiting linear absorption to evaluate the modifications made with respect to the earlier model,\cite{Hestand_15a} in particular focusing on the effect of added triplet states, and to characterize in detail the excited states underlying the absorption band.

Within the framework of Fermi's Golden Rule, the $j$-polarized linear absorption spectrum is expressed as
\begin{align}
A_j(\omega)=\sum_\alpha\vert\Bra{\T{S}_0}\Op{M}_j\Ket{\alpha}\vert^2\;W_j(\omega_\alpha-\omega).
\label{Eq_LinAbs}
\end{align}
Here, $\Op{M}_j$ is the $j$-component of the transition dipole moment operator $\Op{\V{M}}$ of the fission material given by
\begin{align}
\Op{\V{M}}=\sum_m\V{\mu}_m\Ket{\BS{\T{s}_0}_m}\Bra{\BS{\T{s}_1}_m}+\T{H.c.},
\end{align}
where $\V{\mu}_m$ is the transition dipole moment associated with the s$_0\rightarrow\T{s}_1$ optical transition at molecule $m$. Note that molecular transition dipole moments are three-dimensional vector objects (and thus so is the transition dipole moment operator) which are indicated using bold characters to distinguish them from scalar objects. For pentacene, the molecular transition dipole moments are oriented along the short molecular axes,\cite{Smith_10a, Anger_12a} and the applied vector components of $\V{\mu}_m$ are accordingly extracted from the crystal structure.\cite{Holmes_99a} The transition dipole moment operator $\Op{\V{M}}$ acts exclusively on the electronic subspace, effectively inducing a ``vertical'' transition, and as such mixes vibrational states through vibrational overlap factors.\cite{Spano_10a} Also appearing in Eq.~\ref{Eq_LinAbs} are the adiabatic excited states $\alpha$ and associated energies $\omega_\alpha$. These states are coupled through the transition dipole moment operator to the ground state S$_0$ for which all molecules reside in the electronically and vibrationally unexcited state $\Ket{\BS{\T{s}_0}_m,\nu=0}$. And lastly, $W_j(\omega)$ represents the $j$-polarized lineshape function.

\begin{figure}
\includegraphics{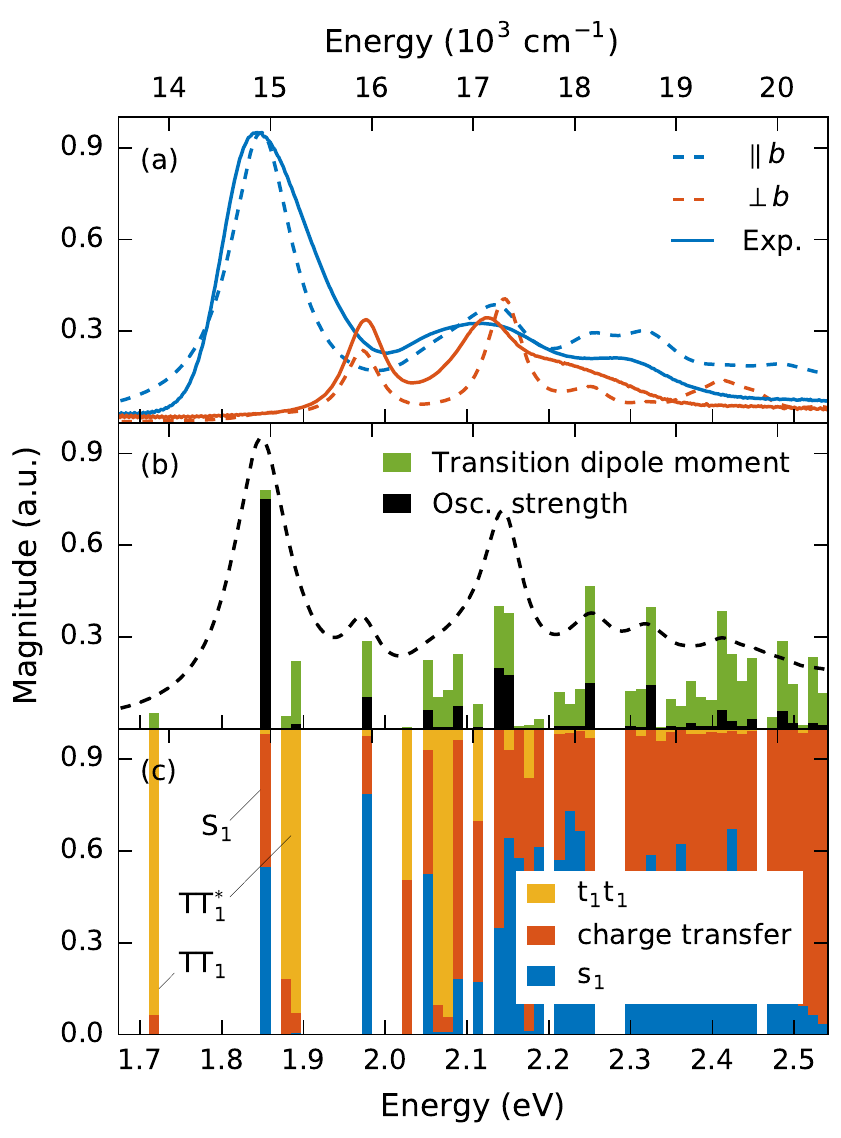}
\caption{(a) Polarization-resolved linear absorption measured for pentacene single crystals (solid curves),\cite{Hestand_15a} shown together with the numerical results from our model for a 6$\times$6 pentacene crystal (dashes). (b) Calculated unpolarized linear absorption (dash). Also shown is a stick spectrum representing the transition dipole moments (green) and oscillator strengths (black) associated with the underlying optical transitions. (c) Dissection of the stick spectrum into diabatic singlet (blue), charge transfer (red), and triplet pair contributions (yellow).}
\label{Fig_LinAbs.pdf}
\end{figure}

Polarized linear absorption of single-crystalline pentacene is calculated through Eq.~\ref{Eq_LinAbs} using a Gaussian lineshape $W_j(\omega)\propto\exp(-\omega^2/2\sigma_j^2)$. In Ref.~\citenum{Hestand_15a}, a polarization-dependent line width $\sigma_j$ was applied to match the measurements, although the physical origin of this property was left undiscussed. Here, we adjust the line widths accordingly in order to unambiguously evaluate the modifications made in our modeling, setting $\sigma_j=46$~meV (374~cm$^{-1}$) for polarization $j$ along the crystallographic $b$-axis, and $\sigma_j=27$~meV (214~cm$^{-1}$) otherwise.\cite{Hestand_15a} Shown in Fig.~\ref{Fig_LinAbs.pdf}(a) are spectra calculated for a 6$\times$6 crystal (72 molecules, 24408 basis states), polarized parallel to $b$ with its perpendicular component in the $ab$-plane, alongside the experimental equivalents recorded for single-crystalline pentacene.\cite{Hestand_15a} When compared to the results reported in Ref.~\citenum{Hestand_15a}, we observe that an excellent agreement is retained upon the modifications made to the modeling. In particular, the Davydov-splitting (energy difference between the lowest-energy peaks in the $\parallel b$ and $\perp b$ spectra) is reproduced, and most of the band features are accounted for. We note that a damping of transitions above 2.3~eV was applied in Ref.~\citenum{Hestand_15a}, representing rapid relaxation resulting from an enhanced density of states at higher energies, which further improves the agreement.\cite{Hestand_15a} We will nevertheless omit such empirical refinements for now, in order not to further complicate the model, although we plan to address the effects of relaxation in a future study. Finally, we have performed comparative calculations excluding the triplet pair excitations from the basis set (see supplementary material), which demonstrate that the impact of these states on linear absorption is generally negligible, with a weak imprint only observable for the second peak polarized perpendicular to $b$.

Since pentacene single-crystals are difficult to grow, most time-resolved spectroscopic measurements on pentacene,\cite{Marciniak_07a, Rao_11a, Wilson_11a, Wilson_13a} including the recent 2DES measurements,\cite{Bakulin_16a} have been performed on polycrystalline samples instead. In what follows, we mimic polycrystallinity by considering unpolarized linear absorption, consisting of the sum of the aforementioned $\parallel b$ and $\perp b$ components plus the component perpendicular to the $ab$-plane, which corresponds to a hypothetical sample of isotropic crystal fragments. Actual polycrystalline pentacene is isotropic only in the $ab$-plane.\cite{Ruiz_04a, Nickel_04a} However, by showing that unpolarized calculations form a good proxy for this situation we simplify the 2DES simulations presented in a follow-up article. The calculated unpolarized linear absorption spectrum is shown in Fig.~\ref{Fig_LinAbs.pdf}(b). The spectrum closely resembles reported measurements of polycrystalline pentacene,\cite{Sebastian_81a, Rao_10a} indicating that our procedure for polycrystallinity indeed appears to be valid for this material. The reason for this is simply that the molecular transition dipole moments of pentacene, being oriented along the short molecular axes, lie largely in the $ab$-plane (see Fig.~\ref{Fig_Crystal}), so that in our approach the isotropy is always projected on this plane, just as is the case for polycrystals.

Also shown in Fig.~\ref{Fig_LinAbs.pdf} is a characterization of the excited states underlying the absorption band. Panel (b) presents a stick spectrum, representing a histogram of the oscillator strengths and transition dipole moments underlying the absorption band, which are defined as $\sum_j\vert\Bra{\T{S}_0}\Op{M}_j\Ket{\alpha}\vert^2$ and its square root, respectively. In an accompanying histogram shown in Fig.~\ref{Fig_LinAbs.pdf}(c), each underlying state $\Ket{\alpha}$ is dissected into contributions from diabatic singlet states (s$_1$), CT states, and triplet pair states (t$_1$t$_1$). One aspect that these data reveal is that the direct mixing between s$_1$ and t$_1$t$_1$ is minimal, albeit non-zero. As a consequence, a direct excitation of triplets through strong mixing into S$_1$, as proposed in Ref.~\citenum{Chan_11a}, seems unlikely based on our data. CT states clearly act as a facilitator of the interaction between singlet and triplets, having significant admixtures in both TT$_1$ and S$_1$. (No significant additional mixing results from the direct two-electron diabatic coupling, the effect of which we have found to be minimal, and which is therefore ignored here.) In the low-energy region, most of the oscillator strength is concentrated in two Davydov components. Consistent with earlier reports,\cite{Yamagata_11a, Beljonne_13a, Berkelbach_14a, Hestand_15a} the lower component at 1.85~eV (14\;900~cm$^{-1}$) is found to consist of an almost 50\%-50\% admixture of s$_1$ and CT states. The upper component at 1.97~eV (15\;900~cm$^{-1}$), on the other hand, has a more pronounced s$_1$ contribution, whereas CT states dominate overall at higher energies.\cite{Hestand_15a} In contrast, we find most of t$_1$t$_1$ to be concentrated in a limited number of states with a small admixture of CT contributions and very small oscillator strengths owing to intensity borrowing from s$_1$. By adjusting the diabatic triplet pair energy E$_{\T{t}_1\T{t}_1}$, we have located the lowest-energy candidates for such triplet-dominated states at 1.72~eV (13\;860~cm$^{-1}$), which is consistent with the direct observation of the fission product states through 2DES,\cite{Bakulin_16a} as is further discussed in Sec.~\ref{Sec_RedShift}.

In order to connect these results to experimental singlet fission studies, we have labeled the involved adiabatic states in Fig.~\ref{Fig_LinAbs.pdf}(c). The lower Davydov component is obviously the strongest absorbing state, and can therefore be regarded as the initially populated exciton.\cite{Smith_10a} This state is thus labeled as S$_1$, which is also consistent with its substantial s$_1$ admixture. The lowest-energy triplet-dominated stick is found to represent a handful of states, which in a corresponding fashion are identified with TT$_1$. Interestingly, an analogous set of states is found roughly 0.17~eV (1380~cm$^{-1}$) higher in energy, slightly above S$_1$, which are found to consist largely of t$_1$t$_1$ contributions dressed with a single vibration $\omega_0$. Using an asterisk to loosely refer to this vibrational sublevel, these states are denoted as TT$_1^*$. A similar vibrationally dressed state was found in the phenomenological modeling supporting the 2DES measurements on pentacene in Ref.~\citenum{Bakulin_16a}, where it was hypothesized that its quasi-resonance with S$_1$ is a key factor in the fission dynamics of this material. TT$_1^*$ also plays an important role in the direct detection of TT$_1$, which we further investigate in a companion article.

Although the low-energy region of the stick spectrum looks quite similar to the one obtained through the phenomenological model reported in Ref.~\citenum{Bakulin_16a}, the oscillator strength of TT$_1$ is found to be about a factor of 5 smaller. This discrepancy may be explained by noting that the phenomenological model does not include CT states, but instead considers two states, s$_1$ and t$_1$t$_1$, coupled directly with an interaction strength of 31~meV (248~cm$^{-1}$). This value is generous compared to the effective interaction strength of 10~meV (81~cm$^{-1}$) derived using perturbation theory including CT intermediates,\cite{Berkelbach_13b} and as such expectedly overestimates the intensity borrowing of TT$_1$ from singlet excitons. Even so, it should be noted that the quantitative accuracy of the perturbative approximation is questionable,\cite{Berkelbach_13b} and that a more reliable measure of the degree of intensity borrowing is obtained through non-perturbative calculations, such as presented here. In order to provide further evidence that our model provides reasonable quantitative estimates for the TT$_1$ oscillator strength, we have performed additional calculations while conforming the electronic basis as well as the parameters to those employed in Ref.~\citenum{Bakulin_16a} (resulting linear absorption and stick spectra are included in the supplementary material). The stick spectra obtained upon these modifications are consistent with the results reported in that work. However, in the corresponding linear absorption spectrum, a distinct feature associated with TT$_1$ shows up as a shoulder of the main S$_1$ absorption peak which is not observed experimentally.\cite{Sebastian_81a, Rao_10a, Hestand_15a} Furthermore, considering the difficulty encountered when experimentally probing the TT$_1$ state, we conclude that the oscillator strength found here is most likely more accurate.

\subsection{Diabatic vs.~adiabatic energies and size convergence}\label{Sec_RedShift}

The Davydov splitting observed for pentacene is a manifestation of the general principle that adiabatic energies $\omega_\alpha$ of coupled molecular systems deviate from the diabatic energies such as $E_{\T{s}_1}$. This deviation is crystal size dependent, and converges with increasing dimension. However, with increasing dimension the vibronic basis employed in our model grows very rapidly. It is therefore crucial to determine the minimal crystal size sufficient to obtain reliable results. This is particularly important in the (expensive) modeling of 2DES.

\begin{figure}
\includegraphics{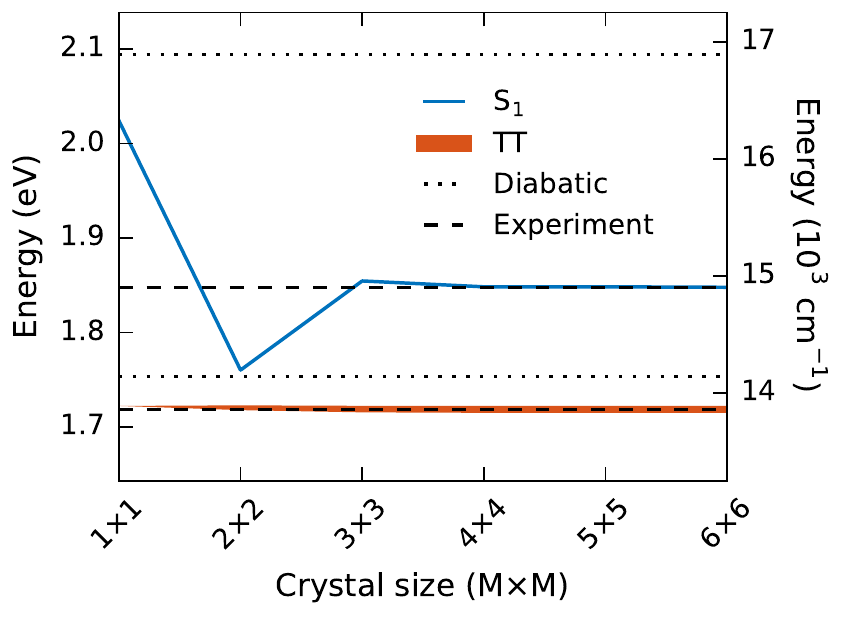}
\caption{Adiabatic S$_1$ (blue curve) and TT$_1$ (red shaded area) energies as a function of the crystal size $M\times M$. Also shown are the experimentally determined adiabatic energies (dashes),\cite{Hestand_15a, Bakulin_16a} and the diabatic energies $E_{\T{s}_1}$ and $E_{\T{t}_1\T{t}_1}$ applied in our calculations (dotted lines at 2.09~eV and 1.75~eV, respectively).}
\label{Fig_RedShift}
\end{figure}

In Fig.~\ref{Fig_RedShift}, we have plotted the adiabatic S$_1$ and TT$_1$ energies calculated for pentacene while varying the crystal size $M\times M$. Also shown are the (size-independent) diabatic energies $E_{\T{s}_1}$ and $E_{\T{t}_1\T{t}_1}$. This figure demonstrates that the adiabatic S$_1$ energy undergoes a dramatic red-shift relative to the diabatic analogue, eventually stabilizing at a value roughly 250~meV (2015~cm$^{-1}$) below $E_{\T{s}_1}$. In marked contrast, the energy of TT$_1$ is apparently fairly insensitive to $M$ and does not change much apart from an initial redshift of about 30~meV (250~cm$^{-1}$). From these data, the minimal crystal size is dictated by the point of convergence of S$_1$, which is found to be at $M=3$. A further investigation (presented in the supplementary material) demonstrates that all of the photophysics in the range up to 2.2~eV is well accounted for by limiting the crystal to this size. In our follow-up article on 2DES we therefore adjust $M$ accordingly.

The adiabatic energy separation between the S$_1$ and TT$_1$ is considered a primary factor impacting singlet fission.\cite{Smith_10a} Nevertheless, values assumed for this separation in pentacene vary from 200~meV (Ref.~\citenum{Smith_10a}) to full resonance.\cite{Chan_11a} This lack of consensus arises to a great extent from the very weak optical transition between TT$_1$ and the (singlet) ground state, which makes the correlated triplet pair energy extremely challenging to detect. Only very recently has this optical transition directly been determined by means of 2DES.\cite{Bakulin_16a} In our modeling, the diabatic energies are adjusted so that the converged energy of TT$_1$ matches this experimentally determined value, similar to the approach used to align the energy of S$_1$ to the experimental value from linear absorption.\cite{Hestand_15a} Importantly, our results show that although the splitting between the measurable adiabatic states is 130~meV, the actual splitting between the diabatic energies $E_{\T{s}_1}$ and $E_{\T{t}_1\T{t}_1}$ is found to be 340~meV, as evinced by Fig.~\ref{Fig_RedShift}. Such a determination of diabatic energies is key to obtaining reliable results from dynamical calculations using a microscopic diabatic basis, especially with regards to the recent indications that resonances between vibrational modes and electronic splittings facilitate rapid singlet fission.\cite{Bakulin_16a, Monahan_16a} In this respect, it is interesting to note that a different diabatic splitting of 170~meV was applied in Ref.~\citenum{Beljonne_13a} to yield a near-resonance between the adiabatic states.\cite{Beljonne_13a} A slightly larger diabatic splitting of 200~meV was employed in the dynamic study by Berkelbach \textit{et al.},\cite{Berkelbach_14a} however, the weaker couplings employed in that work mitigate the red-shift of S$_1$, resulting in a substantial energy offset from TT$_1$. Given these quantitative differences in energy splittings and couplings, it is of great interest to dynamically evaluate singlet fission using our model, which will be the topic of a later study.

\subsection{Characterizing the correlated triplet pair state}\label{Sec_TripletPair}

In addition to its energetic offset from S$_1$, other aspects of the correlated triplet pair state are of importance to singlet fission dynamics. In particular, the degree of spatial separation between the two triplets constituting TT$_1$ expectedly has a profound influence on the dissociation of this intermediate into free triplet excitons.\cite{Pensack_16a} Conventionally, these constituents are envisioned to be located at adjacent molecules, owing to the short range of the charge overlap integrals that couple between S$_1$ and TT$_1$.\cite{Smith_10a, Smith_13a} However, recent magnetic field measurements on tetracene give indications of an inter-triplet separation of two to four times the intermolecular separation.\cite{Wang_15a} Relatively little experimental data is available to elaborate on these contrasting viewpoint, in part due to the dark nature of $TT_1$. Nonetheless, through the microscopic basis of our calculations, information about the spatial anatomy of the correlated triplet pair state in pentacene is readily available.

As a quantitative measure of the spatial composition of TT$_1$, we consider the triplet-triplet correlation operator defined as
\begin{align}
\Op{n}_{\T{t}_1\T{t}_1}(\V{l})=\frac{1}{2}\!\sum_{m\neq m'}\!\!\Ket{\BS{\T{t}_1}_m,\BS{\T{t}_1}_{m'}}\Bra{\BS{\T{t}_1}_m,\BS{\T{t}_1}_{m'}}\;\delta_{\V{l},\V{l}^{m'}-\V{l}^m}.
\label{Eq_TT}
\end{align}
Here, the summation extends over all triplet pairs located at molecules $m$ and $m'$, and the prefactor corrects for double counting. The vectors $\V{l}$, $\V{l}^m$, and $\V{l}^{m'}$ refer to molecular positions in terms of the crystallographic lattice vectors.

\begin{figure}
\includegraphics{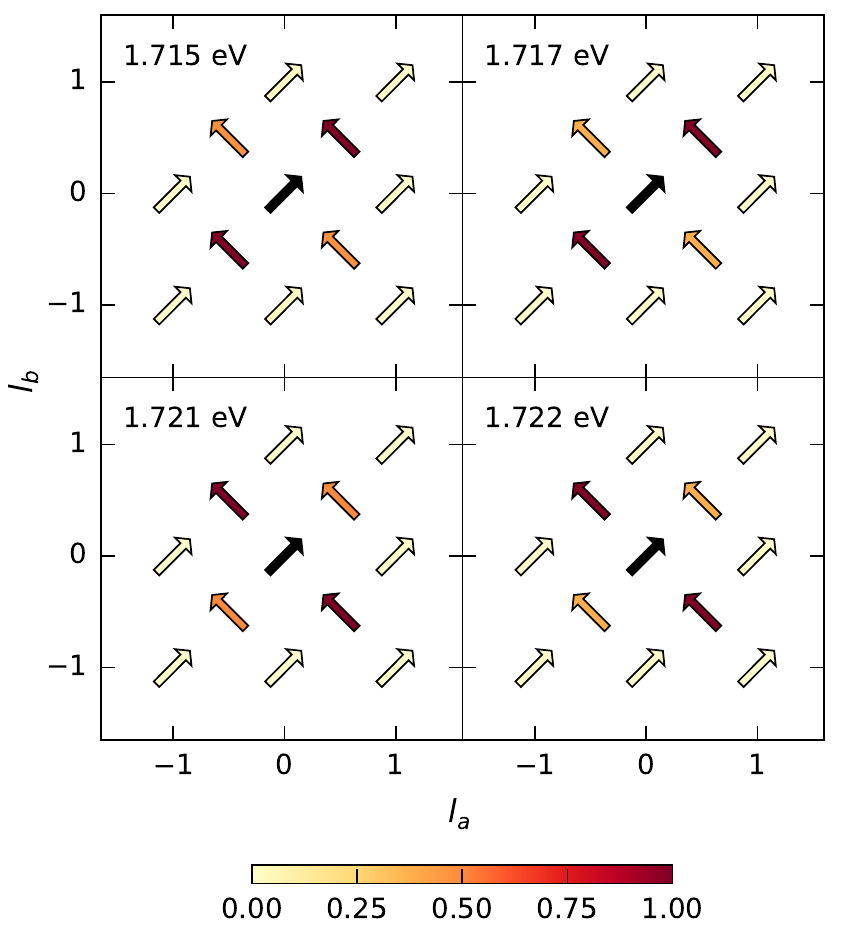}
\caption{Triplet-triplet correlations of the adiabatic states represented by $\T{TT}_1$, calculated for a 3$\times$3 pentacene crystal (associated adiabatic energies shown in top-left corner). For each state, the expectation value of Eq.~\ref{Eq_TT} is shown as a function of $\V{l}=(l_a,l_b)$ by means of colored arrows (loosely referring to the herringbone structure formed by the short molecular axes in the $ab$-plane). The results are normalized individually in each panel for the ease of demonstration. Note that the correlation function is undefined for $l_a=l_b=0$ (corresponding arrow shown as black).}
\label{Fig_TT}
\end{figure}

In our evaluation of the triplet-triplet correlation function, we did not truncate the basis with respect the triplet-triplet and charge separation in order to avoid artificially confined triplet pair states. As this results in a rapidly growing basis set with increasing crystal size, we restricted ourselves to a 3$\times$3 pentacene crystal. For these parameters, TT$_1$ is found to be represented by four near-degenerate triplet-dominated adiabatic states. Fig.~\ref{Fig_TT} shows the triplet-triplet correlation function for each state. From this figure, it is evident that triplets reside predominantly on neighboring molecules in either the $(l_a,l_b)=(1/2,1/2)$ or the $(-1/2,1/2)$ direction, while some mixing between these two configurations is observable. Importantly, the contribution from triplet pairs separated beyond these nearest-neighbor distances is negligible. As such, these findings justify the triplet-triplet truncation radius of 6.2~\AA~applied in the remainder of our calculations. More importantly, however, they confirm the conventional idea of TT$_1$ being largely composed of adjacent triplets. Still, the possibility remains that an enhanced separation is induced by the two-electron couplings that transfer triplet excitons between neighboring molecules in accordance with the well-known Dexter energy transfer mechanism. We have explored this possibility by repeating our calculations while applying typical values for such couplings. The results (shown in the supplementary material) show no quantitative differences with respect to Fig.~\ref{Fig_TT}, so that this idea can safely be discarded. For tetracene, whose difference to pentacene lies mostly in the energetic location of the S$_1$ state, we anticipate a qualitatively similar composition for TT$_1$. We therefore hypothesize that the magnetic field measurements reported in Ref.~\citenum{Wang_15a} relate to the subsequent step in the singlet fission process, in which TT$_1$ is in the process of dissociating into spatially separated triplets. The initial stage of this dissociation, and the potential role of Dexter transfer integrals as well as the effective binding energy between triplets,\cite{Feng_13a, Zeng_14a} will be a topic of interest in our future work.

\section{Conclusions}\label{Sec_Conclusions}

In summary, this article presents a vibronic exciton theory of fission materials which combines a microscopic set of electronic degrees of freedom with an explicit quantum-mechanical treatment of an intramolecular vibrational mode. As such, this article forms a starting point of our ongoing pursuit to unravel the mechanistic principles of singlet fission, in particular addressing the recent experimental indications of the functional relevance of vibronic coupling.\cite{Musser_15a, Bakulin_16a, Monahan_16a} The vibronic model is partly derived from a recent parametrization of pentacene single crystals through polarized linear absorption,\cite{Hestand_15a} which we extended to include the triplet excitons constituting the correlated triplet pair state TT$_1$ through which singlet fission proceeds. In doing so, the associated energy is adjusted in accordance with the recent detection of TT$_1$ by means of 2DES.\cite{Bakulin_16a} We examined the excited states constituting the absorption band of pentacene, paying particular attention to the dependence of adiabatic energies on the applied crystal sizes. Besides providing the diabatic transition energies required to reliably simulate excited state dynamics, our findings revealed that a crystal consisting of $3\times3$ unit cells in the $ab$-plane suffices to account for the photophysics of pentacene in the spectral region of S$_1$ and TT$_1$. Lastly, we provided an insight into the spatial configuration of TT$_1$ demonstrating that the constituent triplets are predominantly located at adjacent molecules, which contradicts a recent experimental report predicting a triplet-triplet separation of several intermolecular distances.\cite{Wang_15a}

In the next article in this series, we focus on the observation of singlet fission through time-resolved spectroscopic experiments, and in particular the recent direct observation of the correlated triplet pair state through 2DES measurements on pentacene.\cite{Bakulin_16a} To this end, the vibronic exciton model introduced in the present article is employed to simulate and analyze 2DES. Our investigation of the crystal size convergence is instrumental to realize such simulations at manageable computational costs. This holds perhaps even more so for the dynamical calculations through which we plan to identify the key factors making successful fission materials, to be presented in future work.

\section*{Supplementary material}

See supplementary material for the electronic couplings and Coulomb energies of crystalline pentacene applied in our modeling, and for supplementary results.

\section*{Acknowledgements}

The authors thank Nick Hestand, Frank Spano, and Tim Berkelbach for providing us with the electronic couplings and Coulomb energies of crystalline pentacene, and for fruitful discussions. R.T.~acknowledges The Netherlands Organisation for Scientific Research NWO for support through a Rubicon grant. D.R.R.~acknowledges funding from NSF grant no.~CHE--1464802.


\begin{thebibliography}{56}%
\makeatletter
\providecommand \@ifxundefined [1]{%
 \@ifx{#1\undefined}
}%
\providecommand \@ifnum [1]{%
 \ifnum #1\expandafter \@firstoftwo
 \else \expandafter \@secondoftwo
 \fi
}%
\providecommand \@ifx [1]{%
 \ifx #1\expandafter \@firstoftwo
 \else \expandafter \@secondoftwo
 \fi
}%
\providecommand \natexlab [1]{#1}%
\providecommand \enquote  [1]{``#1''}%
\providecommand \bibnamefont  [1]{#1}%
\providecommand \bibfnamefont [1]{#1}%
\providecommand \citenamefont [1]{#1}%
\providecommand \href@noop [0]{\@secondoftwo}%
\providecommand \href [0]{\begingroup \@sanitize@url \@href}%
\providecommand \@href[1]{\@@startlink{#1}\@@href}%
\providecommand \@@href[1]{\endgroup#1\@@endlink}%
\providecommand \@sanitize@url [0]{\catcode `\\12\catcode `\$12\catcode
  `\&12\catcode `\#12\catcode `\^12\catcode `\_12\catcode `\%12\relax}%
\providecommand \@@startlink[1]{}%
\providecommand \@@endlink[0]{}%
\providecommand \url  [0]{\begingroup\@sanitize@url \@url }%
\providecommand \@url [1]{\endgroup\@href {#1}{\urlprefix }}%
\providecommand \urlprefix  [0]{URL }%
\providecommand \Eprint [0]{\href }%
\providecommand \doibase [0]{http://dx.doi.org/}%
\providecommand \selectlanguage [0]{\@gobble}%
\providecommand \bibinfo  [0]{\@secondoftwo}%
\providecommand \bibfield  [0]{\@secondoftwo}%
\providecommand \translation [1]{[#1]}%
\providecommand \BibitemOpen [0]{}%
\providecommand \bibitemStop [0]{}%
\providecommand \bibitemNoStop [0]{.\EOS\space}%
\providecommand \EOS [0]{\spacefactor3000\relax}%
\providecommand \BibitemShut  [1]{\csname bibitem#1\endcsname}%
\let\auto@bib@innerbib\@empty
\bibitem [{\citenamefont {Singh}\ \emph {et~al.}(1965)\citenamefont {Singh},
  \citenamefont {Jones}, \citenamefont {Siebrand}, \citenamefont {Stoicheff},\
  and\ \citenamefont {Schneider}}]{Singh_65}%
  \BibitemOpen
  \bibfield  {author} {\bibinfo {author} {\bibfnamefont {S.}~\bibnamefont
  {Singh}}, \bibinfo {author} {\bibfnamefont {W.~J.}\ \bibnamefont {Jones}},
  \bibinfo {author} {\bibfnamefont {W.}~\bibnamefont {Siebrand}}, \bibinfo
  {author} {\bibfnamefont {B.~P.}\ \bibnamefont {Stoicheff}}, \ and\ \bibinfo
  {author} {\bibfnamefont {W.~G.}\ \bibnamefont {Schneider}},\ }\href
  {http://scitation.aip.org/content/aip/journal/jcp/42/1/10.1063/1.1695695}
  {\bibfield  {journal} {\bibinfo  {journal} {J.~Chem.~Phys.}\ }\textbf
  {\bibinfo {volume} {42}},\ \bibinfo {pages} {330} (\bibinfo {year}
  {1965})}\BibitemShut {NoStop}%
\bibitem [{\citenamefont {Avakian}\ and\ \citenamefont
  {Merrifield}(1968)}]{Avakian_68a}%
  \BibitemOpen
  \bibfield  {author} {\bibinfo {author} {\bibfnamefont {P.}~\bibnamefont
  {Avakian}}\ and\ \bibinfo {author} {\bibfnamefont {R.~E.}\ \bibnamefont
  {Merrifield}},\ }\href {http://dx.doi.org/10.1080/15421406808082930}
  {\bibfield  {journal} {\bibinfo  {journal} {Mol.~Crystals}\ }\textbf
  {\bibinfo {volume} {5}},\ \bibinfo {pages} {37} (\bibinfo {year}
  {1968})}\BibitemShut {NoStop}%
\bibitem [{\citenamefont {Smith}\ and\ \citenamefont
  {Michl}(2010)}]{Smith_10a}%
  \BibitemOpen
  \bibfield  {author} {\bibinfo {author} {\bibfnamefont {M.~B.}\ \bibnamefont
  {Smith}}\ and\ \bibinfo {author} {\bibfnamefont {J.}~\bibnamefont {Michl}},\
  }\href {http://dx.doi.org/10.1021/cr1002613} {\bibfield  {journal} {\bibinfo
  {journal} {Chem. Rev.}\ }\textbf {\bibinfo {volume} {110}},\ \bibinfo {pages}
  {6891} (\bibinfo {year} {2010})}\BibitemShut {NoStop}%
\bibitem [{\citenamefont {Smith}\ and\ \citenamefont
  {Michl}(2013)}]{Smith_13a}%
  \BibitemOpen
  \bibfield  {author} {\bibinfo {author} {\bibfnamefont {M.~B.}\ \bibnamefont
  {Smith}}\ and\ \bibinfo {author} {\bibfnamefont {J.}~\bibnamefont {Michl}},\
  }\href {http://dx.doi.org/10.1146/annurev-physchem-040412-110130} {\bibfield
  {journal} {\bibinfo  {journal} {Annu.~Rev.~Phys.~Chem.}\ }\textbf {\bibinfo
  {volume} {64}},\ \bibinfo {pages} {361} (\bibinfo {year} {2013})}\BibitemShut
  {NoStop}%
\bibitem [{\citenamefont {Hanna}\ and\ \citenamefont
  {Nozik}(2006)}]{Hanna_06a}%
  \BibitemOpen
  \bibfield  {author} {\bibinfo {author} {\bibfnamefont {M.~C.}\ \bibnamefont
  {Hanna}}\ and\ \bibinfo {author} {\bibfnamefont {A.~J.}\ \bibnamefont
  {Nozik}},\ }\href
  {http://scitation.aip.org/content/aip/journal/jap/100/7/10.1063/1.2356795}
  {\bibfield  {journal} {\bibinfo  {journal} {J.~Appl.~Phys.}\ }\textbf
  {\bibinfo {volume} {100}},\ \bibinfo {eid} {074510} (\bibinfo {year}
  {2006})}\BibitemShut {NoStop}%
\bibitem [{\citenamefont {Shockley}\ and\ \citenamefont
  {Queisser}(1961)}]{Shockley_61a}%
  \BibitemOpen
  \bibfield  {author} {\bibinfo {author} {\bibfnamefont {W.}~\bibnamefont
  {Shockley}}\ and\ \bibinfo {author} {\bibfnamefont {H.~J.}\ \bibnamefont
  {Queisser}},\ }\href
  {http://scitation.aip.org/content/aip/journal/jap/32/3/10.1063/1.1736034}
  {\bibfield  {journal} {\bibinfo  {journal} {J.~Appl.~Phys.}\ }\textbf
  {\bibinfo {volume} {32}},\ \bibinfo {pages} {510} (\bibinfo {year}
  {1961})}\BibitemShut {NoStop}%
\bibitem [{\citenamefont {Najafov}\ \emph {et~al.}(2010)\citenamefont
  {Najafov}, \citenamefont {Lee}, \citenamefont {Zhou}, \citenamefont
  {Feldman},\ and\ \citenamefont {Podzorov}}]{Najafov_10a}%
  \BibitemOpen
  \bibfield  {author} {\bibinfo {author} {\bibfnamefont {H.}~\bibnamefont
  {Najafov}}, \bibinfo {author} {\bibfnamefont {B.}~\bibnamefont {Lee}},
  \bibinfo {author} {\bibfnamefont {Q.}~\bibnamefont {Zhou}}, \bibinfo {author}
  {\bibfnamefont {L.~C.}\ \bibnamefont {Feldman}}, \ and\ \bibinfo {author}
  {\bibfnamefont {V.}~\bibnamefont {Podzorov}},\ }\href
  {http://dx.doi.org/10.1038/nmat2872} {\bibfield  {journal} {\bibinfo
  {journal} {Nat.~Mater.}\ }\textbf {\bibinfo {volume} {9}},\ \bibinfo {pages}
  {938} (\bibinfo {year} {2010})}\BibitemShut {NoStop}%
\bibitem [{\citenamefont {Irkhin}\ and\ \citenamefont
  {Biaggio}(2011)}]{Irkhin_11a}%
  \BibitemOpen
  \bibfield  {author} {\bibinfo {author} {\bibfnamefont {P.}~\bibnamefont
  {Irkhin}}\ and\ \bibinfo {author} {\bibfnamefont {I.}~\bibnamefont
  {Biaggio}},\ }\href {http://link.aps.org/doi/10.1103/PhysRevLett.107.017402}
  {\bibfield  {journal} {\bibinfo  {journal} {Phys. Rev. Lett.}\ }\textbf
  {\bibinfo {volume} {107}},\ \bibinfo {pages} {017402} (\bibinfo {year}
  {2011})}\BibitemShut {NoStop}%
\bibitem [{\citenamefont {Jadhav}\ \emph {et~al.}(2011)\citenamefont {Jadhav},
  \citenamefont {Mohanty}, \citenamefont {Sussman}, \citenamefont {Lee},\ and\
  \citenamefont {Baldo}}]{Jadhav_11a}%
  \BibitemOpen
  \bibfield  {author} {\bibinfo {author} {\bibfnamefont {P.~J.}\ \bibnamefont
  {Jadhav}}, \bibinfo {author} {\bibfnamefont {A.}~\bibnamefont {Mohanty}},
  \bibinfo {author} {\bibfnamefont {J.}~\bibnamefont {Sussman}}, \bibinfo
  {author} {\bibfnamefont {J.}~\bibnamefont {Lee}}, \ and\ \bibinfo {author}
  {\bibfnamefont {M.~A.}\ \bibnamefont {Baldo}},\ }\href
  {http://dx.doi.org/10.1021/nl104202j} {\bibfield  {journal} {\bibinfo
  {journal} {Nano Lett.}\ }\textbf {\bibinfo {volume} {11}},\ \bibinfo {pages}
  {1495} (\bibinfo {year} {2011})}\BibitemShut {NoStop}%
\bibitem [{\citenamefont {Ehrler}\ \emph {et~al.}(2012)\citenamefont {Ehrler},
  \citenamefont {Wilson}, \citenamefont {Rao}, \citenamefont {Friend},\ and\
  \citenamefont {Greenham}}]{Ehler_12a}%
  \BibitemOpen
  \bibfield  {author} {\bibinfo {author} {\bibfnamefont {B.}~\bibnamefont
  {Ehrler}}, \bibinfo {author} {\bibfnamefont {M.~W.~B.}\ \bibnamefont
  {Wilson}}, \bibinfo {author} {\bibfnamefont {A.}~\bibnamefont {Rao}},
  \bibinfo {author} {\bibfnamefont {R.~H.}\ \bibnamefont {Friend}}, \ and\
  \bibinfo {author} {\bibfnamefont {N.~C.}\ \bibnamefont {Greenham}},\ }\href
  {http://dx.doi.org/10.1021/nl204297u} {\bibfield  {journal} {\bibinfo
  {journal} {Nano Lett.}\ }\textbf {\bibinfo {volume} {12}},\ \bibinfo {pages}
  {1053} (\bibinfo {year} {2012})}\BibitemShut {NoStop}%
\bibitem [{\citenamefont {Congreve}\ \emph {et~al.}(2013)\citenamefont
  {Congreve}, \citenamefont {Lee}, \citenamefont {Thompson}, \citenamefont
  {Hontz}, \citenamefont {Yost}, \citenamefont {Reusswig}, \citenamefont
  {Bahlke}, \citenamefont {Reineke}, \citenamefont {Van~Voorhis},\ and\
  \citenamefont {Baldo}}]{Congreve_13a}%
  \BibitemOpen
  \bibfield  {author} {\bibinfo {author} {\bibfnamefont {D.~N.}\ \bibnamefont
  {Congreve}}, \bibinfo {author} {\bibfnamefont {J.}~\bibnamefont {Lee}},
  \bibinfo {author} {\bibfnamefont {N.~J.}\ \bibnamefont {Thompson}}, \bibinfo
  {author} {\bibfnamefont {E.}~\bibnamefont {Hontz}}, \bibinfo {author}
  {\bibfnamefont {S.~R.}\ \bibnamefont {Yost}}, \bibinfo {author}
  {\bibfnamefont {P.~D.}\ \bibnamefont {Reusswig}}, \bibinfo {author}
  {\bibfnamefont {M.~E.}\ \bibnamefont {Bahlke}}, \bibinfo {author}
  {\bibfnamefont {S.}~\bibnamefont {Reineke}}, \bibinfo {author} {\bibfnamefont
  {T.}~\bibnamefont {Van~Voorhis}}, \ and\ \bibinfo {author} {\bibfnamefont
  {M.~A.}\ \bibnamefont {Baldo}},\ }\href
  {http://science.sciencemag.org/content/340/6130/334} {\bibfield  {journal}
  {\bibinfo  {journal} {Science}\ }\textbf {\bibinfo {volume} {340}},\ \bibinfo
  {pages} {334} (\bibinfo {year} {2013})}\BibitemShut {NoStop}%
\bibitem [{\citenamefont {Yang}\ \emph {et~al.}(2015)\citenamefont {Yang},
  \citenamefont {Tabachnyk}, \citenamefont {Bayliss}, \citenamefont {B{\"o}hm},
  \citenamefont {Broch}, \citenamefont {Greenham}, \citenamefont {Friend},\
  and\ \citenamefont {Ehrler}}]{Yang_15a}%
  \BibitemOpen
  \bibfield  {author} {\bibinfo {author} {\bibfnamefont {L.}~\bibnamefont
  {Yang}}, \bibinfo {author} {\bibfnamefont {M.}~\bibnamefont {Tabachnyk}},
  \bibinfo {author} {\bibfnamefont {S.~L.}\ \bibnamefont {Bayliss}}, \bibinfo
  {author} {\bibfnamefont {M.~L.}\ \bibnamefont {B{\"o}hm}}, \bibinfo {author}
  {\bibfnamefont {K.}~\bibnamefont {Broch}}, \bibinfo {author} {\bibfnamefont
  {N.~C.}\ \bibnamefont {Greenham}}, \bibinfo {author} {\bibfnamefont {R.~H.}\
  \bibnamefont {Friend}}, \ and\ \bibinfo {author} {\bibfnamefont
  {B.}~\bibnamefont {Ehrler}},\ }\href {http://dx.doi.org/10.1021/nl503650a}
  {\bibfield  {journal} {\bibinfo  {journal} {Nano Lett.}\ }\textbf {\bibinfo
  {volume} {15}},\ \bibinfo {pages} {354} (\bibinfo {year} {2015})}\BibitemShut
  {NoStop}%
\bibitem [{\citenamefont {Wilson}\ \emph {et~al.}(2011)\citenamefont {Wilson},
  \citenamefont {Rao}, \citenamefont {Clark}, \citenamefont {Kumar},
  \citenamefont {Brida}, \citenamefont {Cerullo},\ and\ \citenamefont
  {Friend}}]{Wilson_11a}%
  \BibitemOpen
  \bibfield  {author} {\bibinfo {author} {\bibfnamefont {M.~W.~B.}\
  \bibnamefont {Wilson}}, \bibinfo {author} {\bibfnamefont {A.}~\bibnamefont
  {Rao}}, \bibinfo {author} {\bibfnamefont {J.}~\bibnamefont {Clark}}, \bibinfo
  {author} {\bibfnamefont {R.~S.~S.}\ \bibnamefont {Kumar}}, \bibinfo {author}
  {\bibfnamefont {D.}~\bibnamefont {Brida}}, \bibinfo {author} {\bibfnamefont
  {G.}~\bibnamefont {Cerullo}}, \ and\ \bibinfo {author} {\bibfnamefont
  {R.~H.}\ \bibnamefont {Friend}},\ }\href
  {http://pubs.acs.org/doi/abs/10.1021/ja201688h} {\bibfield  {journal}
  {\bibinfo  {journal} {J. Am. Chem. Soc.}\ }\textbf {\bibinfo {volume}
  {133}},\ \bibinfo {pages} {11830} (\bibinfo {year} {2011})}\BibitemShut
  {NoStop}%
\bibitem [{\citenamefont {Chan}\ \emph {et~al.}(2011)\citenamefont {Chan},
  \citenamefont {Ligges}, \citenamefont {Jailaubekov}, \citenamefont {Kaake},
  \citenamefont {Miaja-Avila},\ and\ \citenamefont {Zhu}}]{Chan_11a}%
  \BibitemOpen
  \bibfield  {author} {\bibinfo {author} {\bibfnamefont {W.-L.}\ \bibnamefont
  {Chan}}, \bibinfo {author} {\bibfnamefont {M.}~\bibnamefont {Ligges}},
  \bibinfo {author} {\bibfnamefont {A.}~\bibnamefont {Jailaubekov}}, \bibinfo
  {author} {\bibfnamefont {L.}~\bibnamefont {Kaake}}, \bibinfo {author}
  {\bibfnamefont {L.}~\bibnamefont {Miaja-Avila}}, \ and\ \bibinfo {author}
  {\bibfnamefont {X.-Y.}\ \bibnamefont {Zhu}},\ }\href
  {http://science.sciencemag.org/content/334/6062/1541} {\bibfield  {journal}
  {\bibinfo  {journal} {Science}\ }\textbf {\bibinfo {volume} {334}},\ \bibinfo
  {pages} {1541} (\bibinfo {year} {2011})}\BibitemShut {NoStop}%
\bibitem [{\citenamefont {Bakulin}\ \emph {et~al.}(2016)\citenamefont
  {Bakulin}, \citenamefont {Morgan}, \citenamefont {Kehoe}, \citenamefont
  {Wilson}, \citenamefont {Chin}, \citenamefont {Zigmantas}, \citenamefont
  {Egorova},\ and\ \citenamefont {Rao}}]{Bakulin_16a}%
  \BibitemOpen
  \bibfield  {author} {\bibinfo {author} {\bibfnamefont {A.~A.}\ \bibnamefont
  {Bakulin}}, \bibinfo {author} {\bibfnamefont {S.~E.}\ \bibnamefont {Morgan}},
  \bibinfo {author} {\bibfnamefont {T.~B.}\ \bibnamefont {Kehoe}}, \bibinfo
  {author} {\bibfnamefont {M.~W.~B.}\ \bibnamefont {Wilson}}, \bibinfo {author}
  {\bibfnamefont {A.~W.}\ \bibnamefont {Chin}}, \bibinfo {author}
  {\bibfnamefont {D.}~\bibnamefont {Zigmantas}}, \bibinfo {author}
  {\bibfnamefont {D.}~\bibnamefont {Egorova}}, \ and\ \bibinfo {author}
  {\bibfnamefont {A.}~\bibnamefont {Rao}},\ }\href
  {http://dx.doi.org/10.1038/nchem.2371} {\bibfield  {journal} {\bibinfo
  {journal} {Nat.~Chem.}\ }\textbf {\bibinfo {volume} {8}},\ \bibinfo {pages}
  {16} (\bibinfo {year} {2016})}\BibitemShut {NoStop}%
\bibitem [{\citenamefont {Monahan}\ \emph {et~al.}(2016)\citenamefont
  {Monahan}, \citenamefont {Sun}, \citenamefont {Tamura}, \citenamefont
  {Williams}, \citenamefont {Xu}, \citenamefont {Zhong}, \citenamefont {Kumar},
  \citenamefont {Nuckolls}, \citenamefont {Harutyunyan}, \citenamefont {Chen},
  \citenamefont {Dai}, \citenamefont {Beljonne}, \citenamefont {Rao},\ and\
  \citenamefont {Zhu}}]{Monahan_16a}%
  \BibitemOpen
  \bibfield  {author} {\bibinfo {author} {\bibfnamefont {N.~R.}\ \bibnamefont
  {Monahan}}, \bibinfo {author} {\bibfnamefont {D.}~\bibnamefont {Sun}},
  \bibinfo {author} {\bibfnamefont {H.}~\bibnamefont {Tamura}}, \bibinfo
  {author} {\bibfnamefont {K.~W.}\ \bibnamefont {Williams}}, \bibinfo {author}
  {\bibfnamefont {B.}~\bibnamefont {Xu}}, \bibinfo {author} {\bibfnamefont
  {Y.}~\bibnamefont {Zhong}}, \bibinfo {author} {\bibfnamefont
  {B.}~\bibnamefont {Kumar}}, \bibinfo {author} {\bibfnamefont
  {C.}~\bibnamefont {Nuckolls}}, \bibinfo {author} {\bibfnamefont {A.~R.}\
  \bibnamefont {Harutyunyan}}, \bibinfo {author} {\bibfnamefont
  {G.}~\bibnamefont {Chen}}, \bibinfo {author} {\bibfnamefont {H.-L.}\
  \bibnamefont {Dai}}, \bibinfo {author} {\bibfnamefont {D.}~\bibnamefont
  {Beljonne}}, \bibinfo {author} {\bibfnamefont {Y.}~\bibnamefont {Rao}}, \
  and\ \bibinfo {author} {\bibfnamefont {X.-Y.}\ \bibnamefont {Zhu}},\ }\href
  {http://dx.doi.org/10.1038/nchem.2665} {\bibfield  {journal} {\bibinfo
  {journal} {Nat.~Chem.}\ }\textbf {\bibinfo {volume} {advance online
  publication}},\  (\bibinfo {year} {2016})}\BibitemShut {NoStop}%
\bibitem [{\citenamefont {Zimmerman}\ \emph {et~al.}(2011)\citenamefont
  {Zimmerman}, \citenamefont {Bell}, \citenamefont {Casanova},\ and\
  \citenamefont {Head-Gordon}}]{Zimmerman_11a}%
  \BibitemOpen
  \bibfield  {author} {\bibinfo {author} {\bibfnamefont {P.~M.}\ \bibnamefont
  {Zimmerman}}, \bibinfo {author} {\bibfnamefont {F.}~\bibnamefont {Bell}},
  \bibinfo {author} {\bibfnamefont {D.}~\bibnamefont {Casanova}}, \ and\
  \bibinfo {author} {\bibfnamefont {M.}~\bibnamefont {Head-Gordon}},\ }\href
  {http://dx.doi.org/10.1021/ja208431r} {\bibfield  {journal} {\bibinfo
  {journal} {J. Am. Chem. Soc.}\ }\textbf {\bibinfo {volume} {133}},\ \bibinfo
  {pages} {19944} (\bibinfo {year} {2011})}\BibitemShut {NoStop}%
\bibitem [{\citenamefont {Feng}, \citenamefont {Luzanov},\ and\ \citenamefont
  {Krylov}(2013)}]{Feng_13a}%
  \BibitemOpen
  \bibfield  {author} {\bibinfo {author} {\bibfnamefont {X.}~\bibnamefont
  {Feng}}, \bibinfo {author} {\bibfnamefont {A.~V.}\ \bibnamefont {Luzanov}}, \
  and\ \bibinfo {author} {\bibfnamefont {A.~I.}\ \bibnamefont {Krylov}},\
  }\href {http://dx.doi.org/10.1021/jz402122m} {\bibfield  {journal} {\bibinfo
  {journal} {J.~Phys.~Chem.~Lett.}\ }\textbf {\bibinfo {volume} {4}},\ \bibinfo
  {pages} {3845} (\bibinfo {year} {2013})}\BibitemShut {NoStop}%
\bibitem [{\citenamefont {Parker}\ \emph {et~al.}(2014)\citenamefont {Parker},
  \citenamefont {Seideman}, \citenamefont {Ratner},\ and\ \citenamefont
  {Shiozaki}}]{Parker_14a}%
  \BibitemOpen
  \bibfield  {author} {\bibinfo {author} {\bibfnamefont {S.~M.}\ \bibnamefont
  {Parker}}, \bibinfo {author} {\bibfnamefont {T.}~\bibnamefont {Seideman}},
  \bibinfo {author} {\bibfnamefont {M.~A.}\ \bibnamefont {Ratner}}, \ and\
  \bibinfo {author} {\bibfnamefont {T.}~\bibnamefont {Shiozaki}},\ }\href
  {http://dx.doi.org/10.1021/jp505082a} {\bibfield  {journal} {\bibinfo
  {journal} {J.~Phys.~Chem.~C}\ }\textbf {\bibinfo {volume} {118}},\ \bibinfo
  {pages} {12700} (\bibinfo {year} {2014})}\BibitemShut {NoStop}%
\bibitem [{\citenamefont {Musser}\ \emph {et~al.}(2015)\citenamefont {Musser},
  \citenamefont {Liebel}, \citenamefont {Schnedermann}, \citenamefont {Wende},
  \citenamefont {Kehoe}, \citenamefont {Rao},\ and\ \citenamefont
  {Kukura}}]{Musser_15a}%
  \BibitemOpen
  \bibfield  {author} {\bibinfo {author} {\bibfnamefont {A.~J.}\ \bibnamefont
  {Musser}}, \bibinfo {author} {\bibfnamefont {M.}~\bibnamefont {Liebel}},
  \bibinfo {author} {\bibfnamefont {C.}~\bibnamefont {Schnedermann}}, \bibinfo
  {author} {\bibfnamefont {T.}~\bibnamefont {Wende}}, \bibinfo {author}
  {\bibfnamefont {T.~B.}\ \bibnamefont {Kehoe}}, \bibinfo {author}
  {\bibfnamefont {A.}~\bibnamefont {Rao}}, \ and\ \bibinfo {author}
  {\bibfnamefont {P.}~\bibnamefont {Kukura}},\ }\href
  {http://dx.doi.org/10.1038/nphys3241} {\bibfield  {journal} {\bibinfo
  {journal} {Nat.~Phys.}\ }\textbf {\bibinfo {volume} {11}},\ \bibinfo {pages}
  {352} (\bibinfo {year} {2015})}\BibitemShut {NoStop}%
\bibitem [{\citenamefont {Zimmerman}, \citenamefont {Zhang},\ and\
  \citenamefont {Musgrave}(2010)}]{Zimmerman_10a}%
  \BibitemOpen
  \bibfield  {author} {\bibinfo {author} {\bibfnamefont {P.~M.}\ \bibnamefont
  {Zimmerman}}, \bibinfo {author} {\bibfnamefont {Z.}~\bibnamefont {Zhang}}, \
  and\ \bibinfo {author} {\bibfnamefont {C.~B.}\ \bibnamefont {Musgrave}},\
  }\href {http://dx.doi.org/10.1038/nchem.694} {\bibfield  {journal} {\bibinfo
  {journal} {Nat.~Chem.}\ }\textbf {\bibinfo {volume} {2}},\ \bibinfo {pages}
  {648} (\bibinfo {year} {2010})}\BibitemShut {NoStop}%
\bibitem [{\citenamefont {Havenith}, \citenamefont {de~Gier},\ and\
  \citenamefont {Broer}(2012)}]{Havenith_12a}%
  \BibitemOpen
  \bibfield  {author} {\bibinfo {author} {\bibfnamefont {R.~W.}\ \bibnamefont
  {Havenith}}, \bibinfo {author} {\bibfnamefont {H.~D.}\ \bibnamefont
  {de~Gier}}, \ and\ \bibinfo {author} {\bibfnamefont {R.}~\bibnamefont
  {Broer}},\ }\href {http://dx.doi.org/10.1080/00268976.2012.695810} {\bibfield
   {journal} {\bibinfo  {journal} {Mol.~Phys.}\ }\textbf {\bibinfo {volume}
  {110}},\ \bibinfo {pages} {2445} (\bibinfo {year} {2012})}\BibitemShut
  {NoStop}%
\bibitem [{\citenamefont {Casanova}(2014)}]{Casanova_14a}%
  \BibitemOpen
  \bibfield  {author} {\bibinfo {author} {\bibfnamefont {D.}~\bibnamefont
  {Casanova}},\ }\href {http://dx.doi.org/10.1021/ct4007635} {\bibfield
  {journal} {\bibinfo  {journal} {J.~Chem.~Theory Comput.}\ }\textbf {\bibinfo
  {volume} {10}},\ \bibinfo {pages} {324} (\bibinfo {year} {2014})}\BibitemShut
  {NoStop}%
\bibitem [{\citenamefont {Yamagata}\ \emph {et~al.}(2011)\citenamefont
  {Yamagata}, \citenamefont {Norton}, \citenamefont {Hontz}, \citenamefont
  {Olivier}, \citenamefont {Beljonne}, \citenamefont {Br{\'e}das},
  \citenamefont {Silbey},\ and\ \citenamefont {Spano}}]{Yamagata_11a}%
  \BibitemOpen
  \bibfield  {author} {\bibinfo {author} {\bibfnamefont {H.}~\bibnamefont
  {Yamagata}}, \bibinfo {author} {\bibfnamefont {J.}~\bibnamefont {Norton}},
  \bibinfo {author} {\bibfnamefont {E.}~\bibnamefont {Hontz}}, \bibinfo
  {author} {\bibfnamefont {Y.}~\bibnamefont {Olivier}}, \bibinfo {author}
  {\bibfnamefont {D.}~\bibnamefont {Beljonne}}, \bibinfo {author}
  {\bibfnamefont {J.~L.}\ \bibnamefont {Br{\'e}das}}, \bibinfo {author}
  {\bibfnamefont {R.~J.}\ \bibnamefont {Silbey}}, \ and\ \bibinfo {author}
  {\bibfnamefont {F.~C.}\ \bibnamefont {Spano}},\ }\href
  {http://scitation.aip.org/content/aip/journal/jcp/134/20/10.1063/1.3590871}
  {\bibfield  {journal} {\bibinfo  {journal} {J.~Chem.~Phys.}\ }\textbf
  {\bibinfo {volume} {134}},\ \bibinfo {eid} {204703} (\bibinfo {year}
  {2011})}\BibitemShut {NoStop}%
\bibitem [{\citenamefont {Teichen}\ and\ \citenamefont
  {Eaves}(2012)}]{Teichen_12a}%
  \BibitemOpen
  \bibfield  {author} {\bibinfo {author} {\bibfnamefont {P.~E.}\ \bibnamefont
  {Teichen}}\ and\ \bibinfo {author} {\bibfnamefont {J.~D.}\ \bibnamefont
  {Eaves}},\ }\href {http://dx.doi.org/10.1021/jp208905k} {\bibfield  {journal}
  {\bibinfo  {journal} {J. Phys. Chem. B}\ }\textbf {\bibinfo {volume} {116}},\
  \bibinfo {pages} {11473} (\bibinfo {year} {2012})}\BibitemShut {NoStop}%
\bibitem [{\citenamefont {Beljonne}\ \emph {et~al.}(2013)\citenamefont
  {Beljonne}, \citenamefont {Yamagata}, \citenamefont {Br\'edas}, \citenamefont
  {Spano},\ and\ \citenamefont {Olivier}}]{Beljonne_13a}%
  \BibitemOpen
  \bibfield  {author} {\bibinfo {author} {\bibfnamefont {D.}~\bibnamefont
  {Beljonne}}, \bibinfo {author} {\bibfnamefont {H.}~\bibnamefont {Yamagata}},
  \bibinfo {author} {\bibfnamefont {J.~L.}\ \bibnamefont {Br\'edas}}, \bibinfo
  {author} {\bibfnamefont {F.~C.}\ \bibnamefont {Spano}}, \ and\ \bibinfo
  {author} {\bibfnamefont {Y.}~\bibnamefont {Olivier}},\ }\href
  {http://link.aps.org/doi/10.1103/PhysRevLett.110.226402} {\bibfield
  {journal} {\bibinfo  {journal} {Phys. Rev. Lett.}\ }\textbf {\bibinfo
  {volume} {110}},\ \bibinfo {pages} {226402} (\bibinfo {year}
  {2013})}\BibitemShut {NoStop}%
\bibitem [{\citenamefont {Berkelbach}, \citenamefont {Hybertsen},\ and\
  \citenamefont {Reichman}(2013{\natexlab{a}})}]{Berkelbach_13a}%
  \BibitemOpen
  \bibfield  {author} {\bibinfo {author} {\bibfnamefont {T.~C.}\ \bibnamefont
  {Berkelbach}}, \bibinfo {author} {\bibfnamefont {M.~S.}\ \bibnamefont
  {Hybertsen}}, \ and\ \bibinfo {author} {\bibfnamefont {D.~R.}\ \bibnamefont
  {Reichman}},\ }\href
  {http://scitation.aip.org/content/aip/journal/jcp/138/11/10.1063/1.4794425}
  {\bibfield  {journal} {\bibinfo  {journal} {J.~Chem.~Phys.}\ }\textbf
  {\bibinfo {volume} {138}},\ \bibinfo {eid} {114102} (\bibinfo {year}
  {2013}{\natexlab{a}})}\BibitemShut {NoStop}%
\bibitem [{\citenamefont {Berkelbach}, \citenamefont {Hybertsen},\ and\
  \citenamefont {Reichman}(2013{\natexlab{b}})}]{Berkelbach_13b}%
  \BibitemOpen
  \bibfield  {author} {\bibinfo {author} {\bibfnamefont {T.~C.}\ \bibnamefont
  {Berkelbach}}, \bibinfo {author} {\bibfnamefont {M.~S.}\ \bibnamefont
  {Hybertsen}}, \ and\ \bibinfo {author} {\bibfnamefont {D.~R.}\ \bibnamefont
  {Reichman}},\ }\href
  {http://scitation.aip.org/content/aip/journal/jcp/138/11/10.1063/1.4794427}
  {\bibfield  {journal} {\bibinfo  {journal} {J.~Chem.~Phys.}\ }\textbf
  {\bibinfo {volume} {138}},\ \bibinfo {eid} {114103} (\bibinfo {year}
  {2013}{\natexlab{b}})}\BibitemShut {NoStop}%
\bibitem [{\citenamefont {Berkelbach}, \citenamefont {Hybertsen},\ and\
  \citenamefont {Reichman}(2014)}]{Berkelbach_14a}%
  \BibitemOpen
  \bibfield  {author} {\bibinfo {author} {\bibfnamefont {T.~C.}\ \bibnamefont
  {Berkelbach}}, \bibinfo {author} {\bibfnamefont {M.~S.}\ \bibnamefont
  {Hybertsen}}, \ and\ \bibinfo {author} {\bibfnamefont {D.~R.}\ \bibnamefont
  {Reichman}},\ }\href
  {http://scitation.aip.org/content/aip/journal/jcp/141/7/10.1063/1.4892793}
  {\bibfield  {journal} {\bibinfo  {journal} {J.~Chem.~Phys.}\ }\textbf
  {\bibinfo {volume} {141}},\ \bibinfo {eid} {074705} (\bibinfo {year}
  {2014})}\BibitemShut {NoStop}%
\bibitem [{\citenamefont {Hestand}\ \emph
  {et~al.}(2015{\natexlab{a}})\citenamefont {Hestand}, \citenamefont
  {Yamagata}, \citenamefont {Xu}, \citenamefont {Sun}, \citenamefont {Zhong},
  \citenamefont {Harutyunyan}, \citenamefont {Chen}, \citenamefont {Dai},
  \citenamefont {Rao},\ and\ \citenamefont {Spano}}]{Hestand_15a}%
  \BibitemOpen
  \bibfield  {author} {\bibinfo {author} {\bibfnamefont {N.~J.}\ \bibnamefont
  {Hestand}}, \bibinfo {author} {\bibfnamefont {H.}~\bibnamefont {Yamagata}},
  \bibinfo {author} {\bibfnamefont {B.}~\bibnamefont {Xu}}, \bibinfo {author}
  {\bibfnamefont {D.}~\bibnamefont {Sun}}, \bibinfo {author} {\bibfnamefont
  {Y.}~\bibnamefont {Zhong}}, \bibinfo {author} {\bibfnamefont {A.~R.}\
  \bibnamefont {Harutyunyan}}, \bibinfo {author} {\bibfnamefont
  {G.}~\bibnamefont {Chen}}, \bibinfo {author} {\bibfnamefont {H.-L.}\
  \bibnamefont {Dai}}, \bibinfo {author} {\bibfnamefont {Y.}~\bibnamefont
  {Rao}}, \ and\ \bibinfo {author} {\bibfnamefont {F.~C.}\ \bibnamefont
  {Spano}},\ }\href {http://dx.doi.org/10.1021/acs.jpcc.5b07163} {\bibfield
  {journal} {\bibinfo  {journal} {J. Phys. Chem. C}\ }\textbf {\bibinfo
  {volume} {119}},\ \bibinfo {pages} {22137} (\bibinfo {year}
  {2015}{\natexlab{a}})}\BibitemShut {NoStop}%
\bibitem [{Note1()}]{Note1}%
  \BibitemOpen
  \bibinfo {note} {After completion of our manuscript and during the
  formulation of our dynamical modeling, we became aware of Y.~Fujihashi,
  L.~Chen, A.~Ishizaki, J.~Wang, and Y.~Zhao, J.~Chem.~Phys.~\protect \textbf
  {146}, 044101 (2017), which non-perturbatively explores the effect of
  high-frequency modes on singlet fission using dynamical calculations based on
  a phenomenological model. A comparison of this work with the dynamics that
  emerge from the microscopic Hamiltonian will be made in the third paper of
  this series.}\BibitemShut {Stop}%
\bibitem [{\citenamefont {Wang}\ \emph {et~al.}(2015)\citenamefont {Wang},
  \citenamefont {Zhang}, \citenamefont {Zhang}, \citenamefont {Liu},
  \citenamefont {Wang},\ and\ \citenamefont {Xiao}}]{Wang_15a}%
  \BibitemOpen
  \bibfield  {author} {\bibinfo {author} {\bibfnamefont {R.}~\bibnamefont
  {Wang}}, \bibinfo {author} {\bibfnamefont {C.}~\bibnamefont {Zhang}},
  \bibinfo {author} {\bibfnamefont {B.}~\bibnamefont {Zhang}}, \bibinfo
  {author} {\bibfnamefont {Y.}~\bibnamefont {Liu}}, \bibinfo {author}
  {\bibfnamefont {X.}~\bibnamefont {Wang}}, \ and\ \bibinfo {author}
  {\bibfnamefont {M.}~\bibnamefont {Xiao}},\ }\href
  {http://dx.doi.org/10.1038/ncomms9602} {\bibfield  {journal} {\bibinfo
  {journal} {Nat.~Commun.}\ }\textbf {\bibinfo {volume} {6}},\ \bibinfo {pages}
  {8602} (\bibinfo {year} {2015})}\BibitemShut {NoStop}%
\bibitem [{\citenamefont {Philpott}(1971)}]{Philpott_71a}%
  \BibitemOpen
  \bibfield  {author} {\bibinfo {author} {\bibfnamefont {M.~R.}\ \bibnamefont
  {Philpott}},\ }\href@noop {} {\bibfield  {journal} {\bibinfo  {journal} {J.
  Chem. Phys.}\ }\textbf {\bibinfo {volume} {55}},\ \bibinfo {pages} {2039}
  (\bibinfo {year} {1971})}\BibitemShut {NoStop}%
\bibitem [{\citenamefont {Spano}(2002)}]{Spano_02a}%
  \BibitemOpen
  \bibfield  {author} {\bibinfo {author} {\bibfnamefont {F.~C.}\ \bibnamefont
  {Spano}},\ }\href
  {http://scitation.aip.org/content/aip/journal/jcp/116/13/10.1063/1.1446034}
  {\bibfield  {journal} {\bibinfo  {journal} {J. Chem. Phys.}\ }\textbf
  {\bibinfo {volume} {116}},\ \bibinfo {pages} {5877} (\bibinfo {year}
  {2002})}\BibitemShut {NoStop}%
\bibitem [{\citenamefont {Spano}\ \emph {et~al.}(2009)\citenamefont {Spano},
  \citenamefont {Clark}, \citenamefont {Silva},\ and\ \citenamefont
  {Friend}}]{Spano_09a}%
  \BibitemOpen
  \bibfield  {author} {\bibinfo {author} {\bibfnamefont {F.~C.}\ \bibnamefont
  {Spano}}, \bibinfo {author} {\bibfnamefont {J.}~\bibnamefont {Clark}},
  \bibinfo {author} {\bibfnamefont {C.}~\bibnamefont {Silva}}, \ and\ \bibinfo
  {author} {\bibfnamefont {R.~H.}\ \bibnamefont {Friend}},\ }\href
  {http://dx.doi.org/10.1063/1.3076079} {\bibfield  {journal} {\bibinfo
  {journal} {J.~Chem.~Phys.}\ }\textbf {\bibinfo {volume} {130}},\ \bibinfo
  {pages} {074904} (\bibinfo {year} {2009})}\BibitemShut {NoStop}%
\bibitem [{\citenamefont {Chan}\ \emph {et~al.}(2013)\citenamefont {Chan},
  \citenamefont {Berkelbach}, \citenamefont {Provorse}, \citenamefont
  {Monahan}, \citenamefont {Tritsch}, \citenamefont {Hybertsen}, \citenamefont
  {Reichman}, \citenamefont {Gao},\ and\ \citenamefont {Zhu}}]{Chan_13a}%
  \BibitemOpen
  \bibfield  {author} {\bibinfo {author} {\bibfnamefont {W.-L.}\ \bibnamefont
  {Chan}}, \bibinfo {author} {\bibfnamefont {T.~C.}\ \bibnamefont
  {Berkelbach}}, \bibinfo {author} {\bibfnamefont {M.~R.}\ \bibnamefont
  {Provorse}}, \bibinfo {author} {\bibfnamefont {N.~R.}\ \bibnamefont
  {Monahan}}, \bibinfo {author} {\bibfnamefont {J.~R.}\ \bibnamefont
  {Tritsch}}, \bibinfo {author} {\bibfnamefont {M.~S.}\ \bibnamefont
  {Hybertsen}}, \bibinfo {author} {\bibfnamefont {D.~R.}\ \bibnamefont
  {Reichman}}, \bibinfo {author} {\bibfnamefont {J.}~\bibnamefont {Gao}}, \
  and\ \bibinfo {author} {\bibfnamefont {X.-Y.}\ \bibnamefont {Zhu}},\ }\href
  {http://dx.doi.org/10.1021/ar300286s} {\bibfield  {journal} {\bibinfo
  {journal} {Acc.~Chem.~Res.}\ }\textbf {\bibinfo {volume} {46}},\ \bibinfo
  {pages} {1321} (\bibinfo {year} {2013})}\BibitemShut {NoStop}%
\bibitem [{\citenamefont {Holmes}\ \emph {et~al.}(1999)\citenamefont {Holmes},
  \citenamefont {Kumaraswamy}, \citenamefont {Matzger},\ and\ \citenamefont
  {Vollhardt}}]{Holmes_99a}%
  \BibitemOpen
  \bibfield  {author} {\bibinfo {author} {\bibfnamefont {D.}~\bibnamefont
  {Holmes}}, \bibinfo {author} {\bibfnamefont {S.}~\bibnamefont {Kumaraswamy}},
  \bibinfo {author} {\bibfnamefont {A.~J.}\ \bibnamefont {Matzger}}, \ and\
  \bibinfo {author} {\bibfnamefont {K.~P.~C.}\ \bibnamefont {Vollhardt}},\
  }\href
  {http://dx.doi.org/10.1002/(SICI)1521-3765(19991105)5:11<3399::AID-CHEM3399>3.0.CO;2-V}
  {\bibfield  {journal} {\bibinfo  {journal} {Chem.~- Eur.~J}\ }\textbf
  {\bibinfo {volume} {5}},\ \bibinfo {pages} {3399} (\bibinfo {year}
  {1999})}\BibitemShut {NoStop}%
\bibitem [{\citenamefont {Mattheus}\ \emph {et~al.}(2001)\citenamefont
  {Mattheus}, \citenamefont {Dros}, \citenamefont {Baas}, \citenamefont
  {Meetsma}, \citenamefont {Boer},\ and\ \citenamefont
  {Palstra}}]{Mattheus_01a}%
  \BibitemOpen
  \bibfield  {author} {\bibinfo {author} {\bibfnamefont {C.~C.}\ \bibnamefont
  {Mattheus}}, \bibinfo {author} {\bibfnamefont {A.~B.}\ \bibnamefont {Dros}},
  \bibinfo {author} {\bibfnamefont {J.}~\bibnamefont {Baas}}, \bibinfo {author}
  {\bibfnamefont {A.}~\bibnamefont {Meetsma}}, \bibinfo {author} {\bibfnamefont
  {J.~L.~d.}\ \bibnamefont {Boer}}, \ and\ \bibinfo {author} {\bibfnamefont
  {T.~T.~M.}\ \bibnamefont {Palstra}},\ }\href
  {https://doi.org/10.1107/S010827010100703X} {\bibfield  {journal} {\bibinfo
  {journal} {Acta Crystallogr.~Sect.~C}\ }\textbf {\bibinfo {volume} {57}},\
  \bibinfo {pages} {939} (\bibinfo {year} {2001})}\BibitemShut {NoStop}%
\bibitem [{\citenamefont {Tsiper}\ and\ \citenamefont
  {Soos}(2003)}]{Tsiper_03a}%
  \BibitemOpen
  \bibfield  {author} {\bibinfo {author} {\bibfnamefont {E.~V.}\ \bibnamefont
  {Tsiper}}\ and\ \bibinfo {author} {\bibfnamefont {Z.~G.}\ \bibnamefont
  {Soos}},\ }\href {http://link.aps.org/doi/10.1103/PhysRevB.68.085301}
  {\bibfield  {journal} {\bibinfo  {journal} {Phys. Rev. B}\ }\textbf {\bibinfo
  {volume} {68}},\ \bibinfo {pages} {085301} (\bibinfo {year}
  {2003})}\BibitemShut {NoStop}%
\bibitem [{\citenamefont {Coropceanu}\ \emph {et~al.}(2007)\citenamefont
  {Coropceanu}, \citenamefont {Cornil}, \citenamefont {da~Silva~Filho},
  \citenamefont {Olivier}, \citenamefont {Silbey},\ and\ \citenamefont
  {Br{\'e}das}}]{Coropceanu_07a}%
  \BibitemOpen
  \bibfield  {author} {\bibinfo {author} {\bibfnamefont {V.}~\bibnamefont
  {Coropceanu}}, \bibinfo {author} {\bibfnamefont {J.}~\bibnamefont {Cornil}},
  \bibinfo {author} {\bibfnamefont {D.~A.}\ \bibnamefont {da~Silva~Filho}},
  \bibinfo {author} {\bibfnamefont {Y.}~\bibnamefont {Olivier}}, \bibinfo
  {author} {\bibfnamefont {R.}~\bibnamefont {Silbey}}, \ and\ \bibinfo {author}
  {\bibfnamefont {J.-L.}\ \bibnamefont {Br{\'e}das}},\ }\href
  {http://dx.doi.org/10.1021/cr050140x} {\bibfield  {journal} {\bibinfo
  {journal} {Chem. Rev.}\ }\textbf {\bibinfo {volume} {107}},\ \bibinfo {pages}
  {926} (\bibinfo {year} {2007})}\BibitemShut {NoStop}%
\bibitem [{\citenamefont {Yamagata}\ \emph {et~al.}(2014)\citenamefont
  {Yamagata}, \citenamefont {Maxwell}, \citenamefont {Fan}, \citenamefont
  {Kittilstved}, \citenamefont {Briseno}, \citenamefont {Barnes},\ and\
  \citenamefont {Spano}}]{Yamagata_14a}%
  \BibitemOpen
  \bibfield  {author} {\bibinfo {author} {\bibfnamefont {H.}~\bibnamefont
  {Yamagata}}, \bibinfo {author} {\bibfnamefont {D.~S.}\ \bibnamefont
  {Maxwell}}, \bibinfo {author} {\bibfnamefont {J.}~\bibnamefont {Fan}},
  \bibinfo {author} {\bibfnamefont {K.~R.}\ \bibnamefont {Kittilstved}},
  \bibinfo {author} {\bibfnamefont {A.~L.}\ \bibnamefont {Briseno}}, \bibinfo
  {author} {\bibfnamefont {M.~D.}\ \bibnamefont {Barnes}}, \ and\ \bibinfo
  {author} {\bibfnamefont {F.~C.}\ \bibnamefont {Spano}},\ }\href@noop {}
  {\bibfield  {journal} {\bibinfo  {journal} {J.~Phys.~Chem.~C}\ }\textbf
  {\bibinfo {volume} {118}},\ \bibinfo {pages} {28842} (\bibinfo {year}
  {2014})}\BibitemShut {NoStop}%
\bibitem [{\citenamefont {Hestand}\ \emph
  {et~al.}(2015{\natexlab{b}})\citenamefont {Hestand}, \citenamefont
  {Tempelaar}, \citenamefont {Knoester}, \citenamefont {Jansen},\ and\
  \citenamefont {Spano}}]{Hestand_15b}%
  \BibitemOpen
  \bibfield  {author} {\bibinfo {author} {\bibfnamefont {N.~J.}\ \bibnamefont
  {Hestand}}, \bibinfo {author} {\bibfnamefont {R.}~\bibnamefont {Tempelaar}},
  \bibinfo {author} {\bibfnamefont {J.}~\bibnamefont {Knoester}}, \bibinfo
  {author} {\bibfnamefont {T.~L.~C.}\ \bibnamefont {Jansen}}, \ and\ \bibinfo
  {author} {\bibfnamefont {F.~C.}\ \bibnamefont {Spano}},\ }\href {\doibase
  10.1103/PhysRevB.91.195315} {\bibfield  {journal} {\bibinfo  {journal}
  {Phys.~Rev.~B}\ }\textbf {\bibinfo {volume} {91}},\ \bibinfo {pages} {195315}
  (\bibinfo {year} {2015}{\natexlab{b}})}\BibitemShut {NoStop}%
\bibitem [{\citenamefont {Ito}, \citenamefont {Nagami},\ and\ \citenamefont
  {Nakano}(2015)}]{Ito_15a}%
  \BibitemOpen
  \bibfield  {author} {\bibinfo {author} {\bibfnamefont {S.}~\bibnamefont
  {Ito}}, \bibinfo {author} {\bibfnamefont {T.}~\bibnamefont {Nagami}}, \ and\
  \bibinfo {author} {\bibfnamefont {M.}~\bibnamefont {Nakano}},\ }\href
  {http://dx.doi.org/10.1021/acs.jpclett.5b02249} {\bibfield  {journal}
  {\bibinfo  {journal} {J. Phys. Chem. Lett.}\ }\textbf {\bibinfo {volume}
  {6}},\ \bibinfo {pages} {4972} (\bibinfo {year} {2015})}\BibitemShut
  {NoStop}%
\bibitem [{\citenamefont {McGlynn}, \citenamefont {Azumi},\ and\ \citenamefont
  {Kasha}(1964)}]{McGlynn_64a}%
  \BibitemOpen
  \bibfield  {author} {\bibinfo {author} {\bibfnamefont {S.~P.}\ \bibnamefont
  {McGlynn}}, \bibinfo {author} {\bibfnamefont {T.}~\bibnamefont {Azumi}}, \
  and\ \bibinfo {author} {\bibfnamefont {M.}~\bibnamefont {Kasha}},\ }\href
  {\doibase 10.1063/1.1725145} {\bibfield  {journal} {\bibinfo  {journal}
  {J.~Chem.~Phys.}\ }\textbf {\bibinfo {volume} {40}},\ \bibinfo {pages} {507}
  (\bibinfo {year} {1964})}\BibitemShut {NoStop}%
\bibitem [{\citenamefont {Tempelaar}\ \emph {et~al.}(2013)\citenamefont
  {Tempelaar}, \citenamefont {Stradomska}, \citenamefont {Knoester},\ and\
  \citenamefont {Spano}}]{Tempelaar_13a}%
  \BibitemOpen
  \bibfield  {author} {\bibinfo {author} {\bibfnamefont {R.}~\bibnamefont
  {Tempelaar}}, \bibinfo {author} {\bibfnamefont {A.}~\bibnamefont
  {Stradomska}}, \bibinfo {author} {\bibfnamefont {J.}~\bibnamefont
  {Knoester}}, \ and\ \bibinfo {author} {\bibfnamefont {F.~C.}\ \bibnamefont
  {Spano}},\ }\href {http://dx.doi.org/10.1021/jp310298n} {\bibfield  {journal}
  {\bibinfo  {journal} {J.~Phys.~Chem.~B}\ }\textbf {\bibinfo {volume} {117}},\
  \bibinfo {pages} {457} (\bibinfo {year} {2013})}\BibitemShut {NoStop}%
\bibitem [{\citenamefont {Anger}\ \emph {et~al.}(2012)\citenamefont {Anger},
  \citenamefont {Oss{\'o}}, \citenamefont {Heinemeyer}, \citenamefont {Broch},
  \citenamefont {Scholz}, \citenamefont {Gerlach},\ and\ \citenamefont
  {Schreiber}}]{Anger_12a}%
  \BibitemOpen
  \bibfield  {author} {\bibinfo {author} {\bibfnamefont {F.}~\bibnamefont
  {Anger}}, \bibinfo {author} {\bibfnamefont {J.~O.}\ \bibnamefont {Oss{\'o}}},
  \bibinfo {author} {\bibfnamefont {U.}~\bibnamefont {Heinemeyer}}, \bibinfo
  {author} {\bibfnamefont {K.}~\bibnamefont {Broch}}, \bibinfo {author}
  {\bibfnamefont {R.}~\bibnamefont {Scholz}}, \bibinfo {author} {\bibfnamefont
  {A.}~\bibnamefont {Gerlach}}, \ and\ \bibinfo {author} {\bibfnamefont
  {F.}~\bibnamefont {Schreiber}},\ }\href
  {http://scitation.aip.org/content/aip/journal/jcp/136/5/10.1063/1.3677839}
  {\bibfield  {journal} {\bibinfo  {journal} {J.~Chem.~Phys.}\ }\textbf
  {\bibinfo {volume} {136}},\ \bibinfo {eid} {054701} (\bibinfo {year}
  {2012})}\BibitemShut {NoStop}%
\bibitem [{\citenamefont {Spano}(2010)}]{Spano_10a}%
  \BibitemOpen
  \bibfield  {author} {\bibinfo {author} {\bibfnamefont {F.~C.}\ \bibnamefont
  {Spano}},\ }\href {http://dx.doi.org/10.1021/ar900233v} {\bibfield  {journal}
  {\bibinfo  {journal} {Acc.~Chem.~Res.}\ }\textbf {\bibinfo {volume} {43}},\
  \bibinfo {pages} {429} (\bibinfo {year} {2010})}\BibitemShut {NoStop}%
\bibitem [{\citenamefont {Marciniak}\ \emph {et~al.}(2007)\citenamefont
  {Marciniak}, \citenamefont {Fiebig}, \citenamefont {Huth}, \citenamefont
  {Schiefer}, \citenamefont {Nickel}, \citenamefont {Selmaier},\ and\
  \citenamefont {Lochbrunner}}]{Marciniak_07a}%
  \BibitemOpen
  \bibfield  {author} {\bibinfo {author} {\bibfnamefont {H.}~\bibnamefont
  {Marciniak}}, \bibinfo {author} {\bibfnamefont {M.}~\bibnamefont {Fiebig}},
  \bibinfo {author} {\bibfnamefont {M.}~\bibnamefont {Huth}}, \bibinfo {author}
  {\bibfnamefont {S.}~\bibnamefont {Schiefer}}, \bibinfo {author}
  {\bibfnamefont {B.}~\bibnamefont {Nickel}}, \bibinfo {author} {\bibfnamefont
  {F.}~\bibnamefont {Selmaier}}, \ and\ \bibinfo {author} {\bibfnamefont
  {S.}~\bibnamefont {Lochbrunner}},\ }\href
  {http://link.aps.org/doi/10.1103/PhysRevLett.99.176402} {\bibfield  {journal}
  {\bibinfo  {journal} {Phys. Rev. Lett.}\ }\textbf {\bibinfo {volume} {99}},\
  \bibinfo {pages} {176402} (\bibinfo {year} {2007})}\BibitemShut {NoStop}%
\bibitem [{\citenamefont {Rao}\ \emph {et~al.}(2011)\citenamefont {Rao},
  \citenamefont {Wilson}, \citenamefont {Albert-Seifried}, \citenamefont
  {Di~Pietro},\ and\ \citenamefont {Friend}}]{Rao_11a}%
  \BibitemOpen
  \bibfield  {author} {\bibinfo {author} {\bibfnamefont {A.}~\bibnamefont
  {Rao}}, \bibinfo {author} {\bibfnamefont {M.~W.~B.}\ \bibnamefont {Wilson}},
  \bibinfo {author} {\bibfnamefont {S.}~\bibnamefont {Albert-Seifried}},
  \bibinfo {author} {\bibfnamefont {R.}~\bibnamefont {Di~Pietro}}, \ and\
  \bibinfo {author} {\bibfnamefont {R.~H.}\ \bibnamefont {Friend}},\ }\href
  {http://link.aps.org/doi/10.1103/PhysRevB.84.195411} {\bibfield  {journal}
  {\bibinfo  {journal} {Phys. Rev. B}\ }\textbf {\bibinfo {volume} {84}},\
  \bibinfo {pages} {195411} (\bibinfo {year} {2011})}\BibitemShut {NoStop}%
\bibitem [{\citenamefont {Wilson}\ \emph {et~al.}(2013)\citenamefont {Wilson},
  \citenamefont {Rao}, \citenamefont {Ehrler},\ and\ \citenamefont
  {Friend}}]{Wilson_13a}%
  \BibitemOpen
  \bibfield  {author} {\bibinfo {author} {\bibfnamefont {M.~W.~B.}\
  \bibnamefont {Wilson}}, \bibinfo {author} {\bibfnamefont {A.}~\bibnamefont
  {Rao}}, \bibinfo {author} {\bibfnamefont {B.}~\bibnamefont {Ehrler}}, \ and\
  \bibinfo {author} {\bibfnamefont {R.~H.}\ \bibnamefont {Friend}},\
  }\href@noop {} {\bibfield  {journal} {\bibinfo  {journal} {Acc. Chem. Res.}\
  }\textbf {\bibinfo {volume} {46}},\ \bibinfo {pages} {1330} (\bibinfo {year}
  {2013})}\BibitemShut {NoStop}%
\bibitem [{\citenamefont {Ruiz}\ \emph {et~al.}(2004)\citenamefont {Ruiz},
  \citenamefont {Choudhary}, \citenamefont {Nickel}, \citenamefont {Toccoli},
  \citenamefont {Chang}, \citenamefont {Mayer}, \citenamefont {Clancy},
  \citenamefont {Blakely}, \citenamefont {Headrick}, \citenamefont {Iannotta},\
  and\ \citenamefont {Malliaras}}]{Ruiz_04a}%
  \BibitemOpen
  \bibfield  {author} {\bibinfo {author} {\bibfnamefont {R.}~\bibnamefont
  {Ruiz}}, \bibinfo {author} {\bibfnamefont {D.}~\bibnamefont {Choudhary}},
  \bibinfo {author} {\bibfnamefont {B.}~\bibnamefont {Nickel}}, \bibinfo
  {author} {\bibfnamefont {T.}~\bibnamefont {Toccoli}}, \bibinfo {author}
  {\bibfnamefont {K.-C.}\ \bibnamefont {Chang}}, \bibinfo {author}
  {\bibfnamefont {A.~C.}\ \bibnamefont {Mayer}}, \bibinfo {author}
  {\bibfnamefont {P.}~\bibnamefont {Clancy}}, \bibinfo {author} {\bibfnamefont
  {J.~M.}\ \bibnamefont {Blakely}}, \bibinfo {author} {\bibfnamefont {R.~L.}\
  \bibnamefont {Headrick}}, \bibinfo {author} {\bibfnamefont {S.}~\bibnamefont
  {Iannotta}}, \ and\ \bibinfo {author} {\bibfnamefont {G.~G.}\ \bibnamefont
  {Malliaras}},\ }\href {http://dx.doi.org/10.1021/cm049563q} {\bibfield
  {journal} {\bibinfo  {journal} {Chem.~Mater.}\ }\textbf {\bibinfo {volume}
  {16}},\ \bibinfo {pages} {4497} (\bibinfo {year} {2004})}\BibitemShut
  {NoStop}%
\bibitem [{\citenamefont {Nickel}\ \emph {et~al.}(2004)\citenamefont {Nickel},
  \citenamefont {Barabash}, \citenamefont {Ruiz}, \citenamefont {Koch},
  \citenamefont {Kahn}, \citenamefont {Feldman}, \citenamefont {Haglund},\ and\
  \citenamefont {Scoles}}]{Nickel_04a}%
  \BibitemOpen
  \bibfield  {author} {\bibinfo {author} {\bibfnamefont {B.}~\bibnamefont
  {Nickel}}, \bibinfo {author} {\bibfnamefont {R.}~\bibnamefont {Barabash}},
  \bibinfo {author} {\bibfnamefont {R.}~\bibnamefont {Ruiz}}, \bibinfo {author}
  {\bibfnamefont {N.}~\bibnamefont {Koch}}, \bibinfo {author} {\bibfnamefont
  {A.}~\bibnamefont {Kahn}}, \bibinfo {author} {\bibfnamefont {L.~C.}\
  \bibnamefont {Feldman}}, \bibinfo {author} {\bibfnamefont {R.~F.}\
  \bibnamefont {Haglund}}, \ and\ \bibinfo {author} {\bibfnamefont
  {G.}~\bibnamefont {Scoles}},\ }\href
  {http://link.aps.org/doi/10.1103/PhysRevB.70.125401} {\bibfield  {journal}
  {\bibinfo  {journal} {Phys. Rev. B}\ }\textbf {\bibinfo {volume} {70}},\
  \bibinfo {pages} {125401} (\bibinfo {year} {2004})}\BibitemShut {NoStop}%
\bibitem [{\citenamefont {Sebastian}, \citenamefont {Weiser},\ and\
  \citenamefont {B{\"a}ssler}(1981)}]{Sebastian_81a}%
  \BibitemOpen
  \bibfield  {author} {\bibinfo {author} {\bibfnamefont {L.}~\bibnamefont
  {Sebastian}}, \bibinfo {author} {\bibfnamefont {G.}~\bibnamefont {Weiser}}, \
  and\ \bibinfo {author} {\bibfnamefont {H.}~\bibnamefont {B{\"a}ssler}},\
  }\href {http://www.sciencedirect.com/science/article/pii/0301010481850550}
  {\bibfield  {journal} {\bibinfo  {journal} {Chem.~Phys.}\ }\textbf {\bibinfo
  {volume} {61}},\ \bibinfo {pages} {125 } (\bibinfo {year}
  {1981})}\BibitemShut {NoStop}%
\bibitem [{\citenamefont {Rao}\ \emph {et~al.}(2010)\citenamefont {Rao},
  \citenamefont {Wilson}, \citenamefont {Hodgkiss}, \citenamefont
  {Albert-Seifried}, \citenamefont {B{\"a}ssler},\ and\ \citenamefont
  {Friend}}]{Rao_10a}%
  \BibitemOpen
  \bibfield  {author} {\bibinfo {author} {\bibfnamefont {A.}~\bibnamefont
  {Rao}}, \bibinfo {author} {\bibfnamefont {M.~W.~B.}\ \bibnamefont {Wilson}},
  \bibinfo {author} {\bibfnamefont {J.~M.}\ \bibnamefont {Hodgkiss}}, \bibinfo
  {author} {\bibfnamefont {S.}~\bibnamefont {Albert-Seifried}}, \bibinfo
  {author} {\bibfnamefont {H.}~\bibnamefont {B{\"a}ssler}}, \ and\ \bibinfo
  {author} {\bibfnamefont {R.~H.}\ \bibnamefont {Friend}},\ }\href
  {http://dx.doi.org/10.1021/ja1042462} {\bibfield  {journal} {\bibinfo
  {journal} {J. Am. Chem. Soc.}\ }\textbf {\bibinfo {volume} {132}},\ \bibinfo
  {pages} {12698} (\bibinfo {year} {2010})}\BibitemShut {NoStop}%
\bibitem [{\citenamefont {Pensack}\ \emph {et~al.}(2016)\citenamefont
  {Pensack}, \citenamefont {Ostroumov}, \citenamefont {Tilley}, \citenamefont
  {Mazza}, \citenamefont {Grieco}, \citenamefont {Thorley}, \citenamefont
  {Asbury}, \citenamefont {Seferos}, \citenamefont {Anthony},\ and\
  \citenamefont {Scholes}}]{Pensack_16a}%
  \BibitemOpen
  \bibfield  {author} {\bibinfo {author} {\bibfnamefont {R.~D.}\ \bibnamefont
  {Pensack}}, \bibinfo {author} {\bibfnamefont {E.~E.}\ \bibnamefont
  {Ostroumov}}, \bibinfo {author} {\bibfnamefont {A.~J.}\ \bibnamefont
  {Tilley}}, \bibinfo {author} {\bibfnamefont {S.}~\bibnamefont {Mazza}},
  \bibinfo {author} {\bibfnamefont {C.}~\bibnamefont {Grieco}}, \bibinfo
  {author} {\bibfnamefont {K.~J.}\ \bibnamefont {Thorley}}, \bibinfo {author}
  {\bibfnamefont {J.~B.}\ \bibnamefont {Asbury}}, \bibinfo {author}
  {\bibfnamefont {D.~S.}\ \bibnamefont {Seferos}}, \bibinfo {author}
  {\bibfnamefont {J.~E.}\ \bibnamefont {Anthony}}, \ and\ \bibinfo {author}
  {\bibfnamefont {G.~D.}\ \bibnamefont {Scholes}},\ }\href
  {http://dx.doi.org/10.1021/acs.jpclett.6b00947} {\bibfield  {journal}
  {\bibinfo  {journal} {J. Phys. Chem. Lett.}\ }\textbf {\bibinfo {volume}
  {7}},\ \bibinfo {pages} {2370} (\bibinfo {year} {2016})}\BibitemShut
  {NoStop}%
\bibitem [{\citenamefont {Zeng}, \citenamefont {Hoffmann},\ and\ \citenamefont
  {Ananth}(2014)}]{Zeng_14a}%
  \BibitemOpen
  \bibfield  {author} {\bibinfo {author} {\bibfnamefont {T.}~\bibnamefont
  {Zeng}}, \bibinfo {author} {\bibfnamefont {R.}~\bibnamefont {Hoffmann}}, \
  and\ \bibinfo {author} {\bibfnamefont {N.}~\bibnamefont {Ananth}},\
  }\href@noop {} {\bibfield  {journal} {\bibinfo  {journal}
  {J.~Am.~Chem.~Soc.}\ }\textbf {\bibinfo {volume} {136}},\ \bibinfo {pages}
  {5755} (\bibinfo {year} {2014})}\BibitemShut {NoStop}%
\end{thebibliography}

%

\end{document}